\documentclass[fleqn,10pt]{wlpeerj}

\usepackage{url}
\usepackage{natbib}
\usepackage{amsmath}
\usepackage{amssymb}
\usepackage{graphicx}
\usepackage{color} 
\usepackage{subfig}

\title{Automatic large-scale classification of bird sounds is strongly improved by unsupervised feature learning}
\author{Dan Stowell and Mark D. Plumbley}
\affil{Centre for Digital Music, Queen Mary University of London}

\begin{abstract}%
Automatic species classification of birds from their sound is a computational tool of increasing importance in ecology, conservation monitoring and vocal communication studies. To make classification useful in practice, it is crucial to improve its accuracy while ensuring that it can run at big data scales.
Many approaches use acoustic measures based on spectrogram-type data, such as the Mel-frequency cepstral coefficient (MFCC) features which represent a manually-designed summary of spectral information. However, recent work in machine learning has demonstrated that features learnt automatically from data can often outperform manually-designed feature transforms. Feature learning can be performed at large scale and ``unsupervised'', meaning it requires no manual data labelling, yet it can improve performance on ``supervised'' tasks such as classification.
In this work we introduce a technique for feature learning from large volumes of bird sound recordings, inspired by techniques that have proven useful in other domains. We experimentally compare twelve different feature representations derived from the Mel spectrum (of which six use this technique), using four large and diverse databases of bird vocalisations, with a random forest classifier.
We demonstrate that MFCCs are of limited power in this context, leading to worse performance than the raw Mel spectral data.
Conversely, we demonstrate that unsupervised feature learning provides a substantial boost over MFCCs and Mel spectra without adding computational complexity after the model has been trained. The boost is particularly notable for single-label classification tasks at large scale.
The spectro-temporal activations learned through our procedure resemble spectro-temporal receptive fields calculated from avian primary auditory forebrain.
However, for one of our datasets, which contains substantial audio data but few annotations, increased performance is not discernible. We study the interaction between dataset characteristics and choice of feature representation through further empirical analysis.
\end{abstract}

\begin{document}

\flushbottom
\maketitle
\thispagestyle{empty}

%%%%%%%%%%%%%%%%%%%%%%%%%%%%%%%%%%%%%%%%%%%%%%%
\section{Introduction}
\label{sec:intro}

Automatic species classification of birds from their sounds has many potential applications in conservation, ecology and archival \citep{Laiolo:2010,Digby:2013,Ranft:2004}.
However, to be useful it must work with high accuracy across large numbers of possible species, on noisy outdoor recordings and at big data scales.
The ability to scale to big data is crucial:
remote monitoring stations can generate huge volumes of audio recordings \citep{Aide:2013},
and audio archives contain large volumes of audio, much of it without detailed labelling.
For example the British Library Sound Archive holds over 100,000 recordings of bird sound in digital format, from various sources \citep{Ranft:2004}.
Big data scales also imply that methods must work without manual intervention,
in particular without manual segmentation of recordings into song syllables, or into vocal/silent sections.
The lack of segmentation is a pertinent issue for both remote monitoring and archive collections,
since many species of bird may be audible for only a minority of the recorded time,
and therefore much of the audio will contain irrelevant information.

% SOME BIRD RECOG WORK HAS BEEN DONE IN THE PAST
The task of classifying bird sounds by species has been studied by various authors, at least as far back as
\cite{McIlraith:1997}. % "OLDEST" species classif? -- 6 species
(See \cite{Stowell:2010e} for a survey.)
Many of the early studies used small datasets, often noise-free and/or manually-segmented
and with a small number of species, so their practical applicability for ecological applications is unclear.
More recent studies have fewer such limitations, and introduce useful methods customised to the task
\citep{Lakshminarayanan:2009}% A Syllable-Level Probabilistic Framework for Bird Species Identification
\citep{Damoulas:2010}%
\citep{Briggs:2012}%  title={Acoustic classification of multiple simultaneous bird species: A multi-instance multi-label approach}
. However, there remain questions of scalability,
due to the computational intensity of algorithms or to procedures such as all-pairs comparisons which cannot be applied to arbitrarily large datasets without modification \citep{Damoulas:2010}.

In addition to noise-robustness and scalability issues, one further issue
is the number of species considered by a classifier:
certain classification systems may be developed to distinguish among ten or twenty species,
but in many parts of the world there are hundreds of species that might be heard \citep{BTO:2013}.
Further, typical recordings in the wild contain sounds from more than one bird, and so it is advantageous to consider the task as a \textit{multi-label} task,
in which the classifier must return not one label but a set of labels representing all species that are present \citep{Briggs:2012}.

% THE SABIOD STUFF HAS DONE WELL AT PUSHING THIS FORWARD
One recent research project (named ``SABIOD'') has provided a valuable stimulus to the research community
by conducting classification challenges evaluated on large datasets of bird sounds collected in the wild,
and with large numbers of species to recognise
\citep{Glotin:2013}% 
\citep{Fodor:2013}% The Ninth Annual {MLSP} Competition: First place
\citep{birdclef2014}%
.
The research reported in this paper benefits from the datasets made available through that project,
as well as other datasets, to evaluate bird sound classification suitable for large-scale practical deployments.

Some previous work has compared the performance of different classification algorithms for the task \citep{Acevedo:2009,Briggs:2009}.
In the present work, we instead use a standard but powerful classification algorithm,
and focus on the choice of audio features used as input data.
We introduce the concept of \textit{feature learning} which has been applied in other machine learning domains,
and show that in most cases it can lead the classifier to strongly outperform those using common MFCC and Mel spectrum features.
We also evaluate the role of other aspects such as noise reduction in the feature preprocessing;
however, the strongest effect of the parameters we study comes from replacing MFCCs with learned features.

In the following, we use four large and diverse birdsong datasets with varying characteristics to evaluate classifier performance.
Overall, feature learning enables a classifier to perform very strongly on large datasets with large numbers of species,
and achieves this boost with very little computational cost after the training step.
Three of the four datasets demonstrate clearly the boost attained through feature learning,
attaining very strong performance in both single-label and multi-label classification tasks.
One dataset, consisting of long dawn-chorus recordings with a substantial amount of audio but few annotations,
does not yield a significant benefit from the improved feature representation.
We explore the reasons for this in follow-up experiments in which the training data is augmented or substituted with other data.
Before describing our experiment, however, we discuss the use of spectral features and feature learning for audio classification.

\subsection{Spectral features and feature learning}
\label{sec:kfl}

Raw audio data is not generally suitable input to a classification algorithm:
even if the audio inputs were constrained to a fixed duration (so that the data dimensionality was constant),
the dimensionality of an audio signal (considered as a vector) would be extremely large,
and would not represent sound in such a way that acoustically/perceputally similar sounds would generally be near neighbours in the vector space.
Hence audio data is usually converted to a spectrogram-like representation before processing,
i.e.\ the magnitudes of short-time Fourier transformed (STFT) frames of audio, often around 10 ms duration per frame.
(Alternatives to STFT which have been considered for bird sound classification include linear prediction \citep{Fox:2008}, wavelets \citep{Selin:2007} and chirplets \citep{Stowell:2014}).
An STFT spectrum indicates the energy present across a linear range of frequencies.
This linear range might not reflect the perceptual range of a listener, and/or the range of frequencies at which the signal carries information content,
so it is common to transform the frequency axis to a more perceptual scale, such as the Mel scale originally intended to represent the approximately logarithmic sensitivity of human hearing.
This also reduces the dimensionality of the spectrum, but even the Mel spectrum has traditionally been considered rather high-dimensional for automatic analysis.
A convention, originating from speech processing, is to transform the Mel spectrum using a cepstral analysis and then to keep the lower coefficients (e.g.\ the first 13) which typically contain most of the energy %
\citep{Davis:1980}% MFCCs general presentation
. These coefficients, the Mel frequency cepstral coefficients (MFCCs), became widespread in applications of machine learning to audio, including bird vocalisations \citep{Stowell:2010e}.

MFCCs have some advantages, including that the feature values are approximately decorrelated from each other,
and they give a substantially dimension-reduced summary of spectral data.
Dimension reduction is advantageous for manual inspection of data,
and also for use in systems that cannot cope with high-dimensional data.
However, as we will see, modern classification algorithms can cope very well with high-dimensional data,
and dimension reduction always reduces the amount of information that can be made available to later processing,
risking discarding information that a classifier could have used.
Further, there is little reason to suspect that MFCCs should capture information optimal for bird species identification:
they were designed to represent human speech, yet humans and birds differ in their use of the spectrum both perceptually and for production.
MFCCs aside, one could use raw (Mel-)spectra as input to a classifier,
or one could design a new transformation of the spectral data that would tailor the representation to the subject matter.
Rather than designing a new representation manually, we consider automatic feature learning.

The topic of feature learning (or representation learning, dictionary learning)
has been considered from many perspectives within the realm of statistical signal processing
\citep{Bengio:2013}%  title={Representation Learning: A Review and New Perspectives},
\citep{Jafari:2011}%  Fast dictionary learning for sparse representations of speech signals
\citep{Coates:2012}
\citep{Dieleman:2013}%  Multiscale approaches to music audio feature learning
. The general aim is for an algorithm to learn some transformation that, when applied to data, improves performance on tasks such as sparse coding, signal compression or classification.
This procedure may be performed in a ``supervised'' manner, meaning it is supplied with data as well as some side information about the downstream task (e.g.\ class labels), or ``unsupervised'', operating on a dataset but with no information about the downstream task.
A simple example that can be considered to be unsupervised feature learning is principal components analysis (PCA):
applied to a dataset, PCA chooses a linear projection which ensures that the dimensions of the transformed data are decorrelated \citep{Bengio:2013}.
It therefore creates a new feature set, without reference to any particular downstream use of the features,
simply operating on the basis of qualities inherent in the data.

Recent work in machine learning has shown that unsupervised feature learning
can lead to representations that perform very strongly in classification tasks,
despite their ignorance of training data labels that may be available \citep{Coates:2012,Bengio:2013}.
This rather surprising outcome suggests that feature learning methods emphasise patterns in the data that turn out to have semantic relevance,
patterns that are not already made explicit in the basic feature processing such as STFT.
A second surprising aspect is that such representations often perform the opposite of feature reduction,
increasing the dimensionality of the problem without adding any new information:
a deterministic transformation from one feature space to a higher-dimensional feature space
cannot, in an information-theoretic sense, add any information that is not present in the original space.
However, such a transformation can help to reveal the manifold structure that may be present in the data
\citep{Olshausen:2004}% Sparse coding of sensory inputs  ---- helps motivate high-dim fl
. Neural networks, both in machine implemetations and in animals,
perform such a dimension expansion in cases where one layer of neurons is connected as input to a larger layer of neurons
\citep{Olshausen:2004}.
%Biological neurons can be characterised by their \textit{receptive fields} \citep{Hausberger:2000},
%which, analogously to learnt features in machine learning, indicate the sensory input patterns to which they respond.

In our study, however, we will not use a feature learning procedure intended to parallel a biological process.
Instead, we use \textit{spherical k-means}, a simple and highly scalable modification of the classic k-means algorithm
\citep{Coates:2012,Dieleman:2013}.
We perform a further adaptation of the algorithm to ensure that it can
run in streaming fashion across large audio datasets, to be described in Section \ref{sec:methods}.

Birdsong often contains rapid temporal modulations, and this information should be useful for identifying species-specific characteristics \citep{Stowell:2014}.
From this perspective, a useful aspect of feature learning is that it can be applied not only to single spectral frames,
but to short sequences (or ``patches'') of a few frames.
The representation can then reflect not only characteristics of instantaneous frequency patterns in the input data, but characteristics of frequencies and their short-term modulations,
such as chirps sweeping upwards or downwards.
This bears some analogy with the ``delta-MFCC'' features sometimes used by taking the first difference in the time series of MFCCs,
but is more flexible since it can represent amplitude modulations, frequency modulations, and correlated modulations of both sorts
(cf.\ \cite{Stowell:2014}).
In our study we tested variants of feature learning with different temporal structures:
either considering one frame at a time (which does not capture modulation),
multiple frames at a time,
or a variant with two layers of feature learning,
which captures modulation across two timescales.

%%%%%%%%%%%%%%%%%%%%%%%%%%%%%%%%%%%%%%%%%%%%%%%
\section{Materials and Methods}
\label{sec:methods}

Our primary experiment evaluated automatic species classification separately across four different datasets of bird sound.
For each dataset we trained and tested a random forest classifier \citep{Breiman:2001},
while systematically varying the following configuration parameters to determine their effect on performance:
\begin{itemize}
	\item	Choice of features (MFCCs, Mel spectra, or learned features) and their summarisation over time (mean and standard deviation, maximum, or modulation coefficients);
	\item	Whether or not to apply noise reduction to audio spectra as a pre-processing step;
	\item	Decision windowing: whether to treat the full-length audio as a single unit for training/testing purposes, or whether to divide it into shorter-duration windows (1, 5 or 60 seconds);
	\item	How to produce an overall decision when using decision windowing (via the mean or the maximum of the probabilities);
	\item	Classifier configuration: the same random forest classifier tested in single-label, multilabel or binary-relevance setting.
\end{itemize}
We will say more about the configuration parameters below.
Each of the above choices was tested in all combinations (a ``grid search'' over possible configurations) for each of our datasets separately,
thus providing a rigorous search over a vast number of classifier settings,
in hundreds of individual crossvalidated classification tests.

In follow-up experiments we explored some further issues and their effect on species recognition:
\begin{itemize}
	\item	We separated out two aspects of our different feature sets---their dimensionality and their intrinsic character---by projecting the feature data to the same fixed dimensionality, and then re-testing with these;
	\item	We tested the effect of data expansion, by training on the union of two datasets;
	\item	We tested the effect of cross-condition training, by training on one dataset and testing with a different dataset.
\end{itemize}

%%%%%%%%%%%%%%%%%%%%%
\subsection{Datasets}

\begin{table}[t]
\resizebox{\textwidth}{!}{
\begin{tabular}{ l | l r r r r l }
	Dataset	&	Location	&	Items	&	Total duration	&	Mean duration	&	Classes	& Labelling \\
	\hline
	nips4b	&      France	&  687       &  0.8 hrs (125k frames) & 4 secs &  87  & multilabel	\\
	xccoverbl &     UK/Europe &  264     &  4.9 hrs (763k frames) & 67 secs &  88  & single-label	\\
	bldawn	&	UK	& 60         & 7.8 hrs (1.2M frames) & 468 secs & 77  & multilabel	\\ 
	lifeclef2014      & Brazil  & 9688   & 77.8 hrs (12M frames) & 29 secs &  501 & single-label
\end{tabular}
}
\caption{Summary of bird sound datasets used.}
\label{tbl:datasets}
\end{table}

We gathered four datasets, each representing a large amount of audio data and a large number of species to classify (Table \ref{tbl:datasets}).
Two of the datasets (\textit{nips4b} and \textit{lifeclef2014}) consist of the publicly-released training data
from bird classification challenges organised by the SABIOD project \citep{Glotin:2013,birdclef2014}.
The \textit{nips4b} dataset is multilabel (median 1 species per recording, range 0--6);
the \textit{lifeclef2014} dataset is single-label but much larger.
Note that we only use the publicly-released training data from those challenges,
and not any private test data,
and so our evaluation will be similar in nature to their final results but not precisely comparable.
For evaluation we partitioned each of these datasets into two, so that we could run two-fold crossvalidation:
training on one half of the dataset and testing on the other half, and vice versa.

In addition, the British Library Sound Archive has a large collection of environmental sound recordings,
and they made available to us a subset of 60 ``dawn chorus'' recordings.
This consisted of 20 recordings each from three UK-based recordists,
ranging in duration from 2 minutes to 20 minutes,
and annotated by each recordist with a list of species heard (median 6 species per recording, range 3--12).
We refer to this dataset as \textit{bldawn}, and perform three-fold stratified crossvalidation:
for each recordist, we train the system using the data from the other two recordists, and then test on the audio from the held-out recordist.
This stratified approach is useful because it tests whether the system can generalise to recordings from unknown recordists,
rather than adapting to any specifics of the known recordists.

We also gathered a single-label dataset as a subset of the recordings available from the Xeno Canto website,%
\footnote{\url{http://www.xeno-canto.org/}}
covering many of the common UK bird species, and covering at least all the species present in the \textit{bldawn} dataset.
For each species included, we queried Xeno Canto to retrieve three different recordings,
preferring to retrieve recordings from the UK, but allowing the system to return recordings from further afield if too few UK recordings were available.
Our search query also requested high-quality recordings (quality label `A'),
and song rather than calls, where possible.
Since we retrieved three examples for each species, this enabled us to partition the dataset for three-fold crossvalidation:
not stratified into individual recordists (as was \textit{bldawn}), but sampled from a wide range of recordists.

These datasets have widely varying characteristics, for example in the typical duration of the sound files, the recording location,
and the number of classes to distinguish (Table \ref{tbl:datasets}).
Note that most of the datasets have different and irreconcilable lists of class labels:
in particular, for \textit{bldawn} and \textit{xccoverbl} the class label is the species,
whereas \textit{nips4b} and \textit{lifeclef2014} use separate labels for song and calls.
Of our datasets only \textit{bldawn} and \textit{xccoverbl} have strong overlap in their species lists.
Therefore only these datasets could be combined to create larger pools of training data.

In this work we performed automatic classification for each audio file, without any segmentation procedure
to select region(s) of bird vocalisation in the file.
The only segmentation that is done is implicit in the collection processes for the dataset:
for the two datasets originating from Xeno Canto, each audio clip might or might not contain a large amount of silence or other noise, depending on the contributor;
for \textit{nips4b} the audio is collected from remote monitoring stations with no manual selection;
for \textit{bldawn} the audio is selected by the contributor, but not trimmed to a specific vocalisation, instead selected to present a long dawn chorus audio recording.

%%%%%%%%%%%%%%%%%%%%%%%%%%%%%%%%%%%%%%%%%%%%%%%%
\subsection{Feature learning method}

\begin{figure}[tp]
	\begin{center}
	\includegraphics [width=0.5\textwidth,clip, trim=40mm 10mm 40mm 10mm]{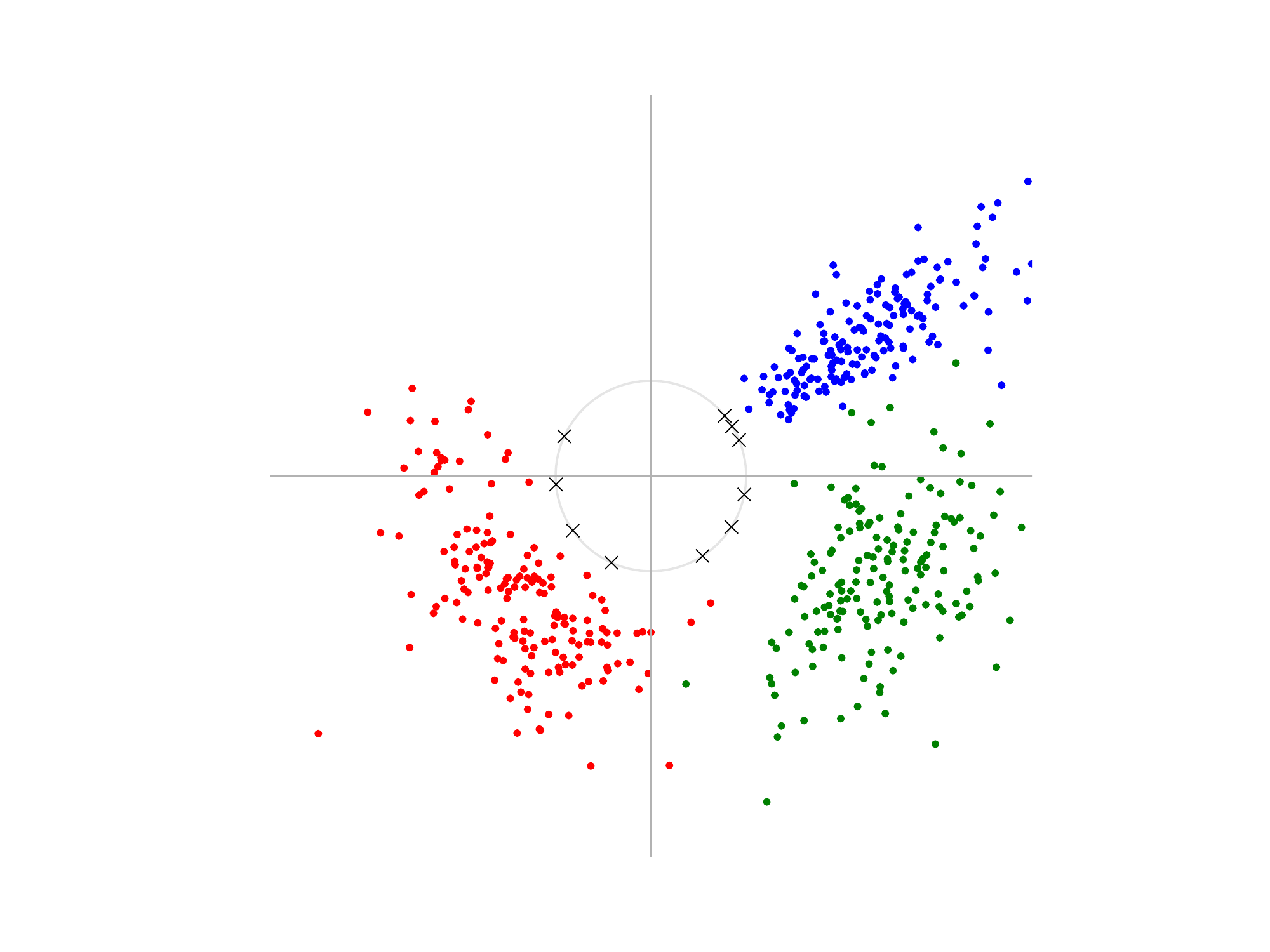}
	\end{center}
	\caption{Example of spherical k-means applied to a simple two-dimensional dataset. We generated synthetic 2D data points by sampling from three clusters which were each Gaussian-distributed in terms of their angle and log-magnitude (coloured dots), and then applied our online spherical k-means algorithm to find 10 unit vectors (crosses). These unit vectors form an overcomplete basis with which one could represent this toy data, projecting two-dimensional space to ten-dimensional space.}
	\label{fig:oskmeansexample}
\end{figure}

As discussed in Section \ref{sec:kfl}, the aim of unsupervised feature learning is to find some transformation of a dataset,
driven only by the characteristics inherent in that dataset.
For this we use a method that has has shown promise in previous studies, and can be run effectively at big data scales:
\textit{spherical k-means}, described by \cite{Coates:2012} and first applied to audio by \cite{Dieleman:2013}.
Spherical k-means is related to the well-known k-means clustering algorithm \citep{Lloyd:1982},
except that instead of searching for cluster centroids which minimise the Euclidean distance to the data points,
we search for unit vectors (directions) to minimise their angular distance from the data points.
This is achieved by modifying the iterative update procedure for the k-means algorithm:
for an input data point, rather than finding the nearest centroid by Euclidean distance and then moving the centroid towards that data point,
the nearest centroid is found by cosine distance,
\begin{equation}
	\text{cosine distance} = 1-\cos(\theta) = 1-{A \cdot B \over \|A\| \|B\|} \, ,
\end{equation}
where $A$ and $B$ are vectors to be compared, $\theta$ is the angle between them, and $\|\cdot\|$ is the Euclidean vector norm.
The centroid is renormalised after update so that it is always a unit vector.
Fig. \ref{fig:oskmeansexample} shows an example of spherical k-means applied to synthetic data.
Spherical k-means thus finds a set of unit vectors which represent the distribution of directions found in the data:
it finds a basis (here an overcomplete basis) so that data points can in general be well approximated as a scalar multiple of one of the basis vectors.
This basis can then be used to represent input data in a new feature space which reflects the discovered regularities,
in the simplest case by representing every input datum by its dot product with each of the basis vectors \citep{Coates:2012,Dieleman:2013}:
\begin{equation}
	x'(n,j) = \sum_{i=1}^M{b_j(i) x(n,i)}    \, ,
	\label{eqn:dotprod1frame}
\end{equation}
where $x$ represents the input data indexed by time frame $n$ and feature index $i$ (with $M$ the number of input features, e.g.\ the number of spectral bins),
$b_j$ is one of the learnt basis vectors (indexed by $j \in [1,k]$),
and $x'$ is the new feature representation.
In our case, the data on which we applied the spherical k-means procedure consisted of Mel spectral frames ($M=40$ dimensions),
which we first normalised and PCA-whitened as in \cite{Dieleman:2013}.

We also tested configurations in which the input data was not one spectral frame but a sequence of them---e.g.\ a sequence of four spectral frames at a time---allowing the clustering to respond to short-term temporal patterns as well as spectral patterns.
We can write this as
\begin{equation}
	x'(n,j) = \sum_{\delta=0}^\Delta{\sum_{i=1}^M{b_j(\delta,i) x(n+\delta,i)}}    \, ,
	\label{eqn:dotprodmultiframe}
\end{equation}
where $\Delta$ is the number of frames considered at a time,
and the $b$ are now indexed by a frame-offset as well as the feature index.
(See Fig. \ref{fig:basesexample} to preview examples of such bases.)
Alternatively, this can be thought of as ``stacking'' frames, e.g.\
stacking each sequence of four 40-dimensional spectral frames to give a 160-dimensional vector,
before applying \eqref{eqn:dotprod1frame} as before.
In all our experiments we used a fixed $k=500$, a value which has been found useful in previous studies \citep{Dieleman:2013}.

The standard implementation of k-means clustering requires an iterative batch process which considers all data points in every step.
This is not feasible for high data volumes.
Some authors use ``minibatch'' updates, i.e.\ subsamples of the dataset.
For scalability as well as for the potential to handle real-time streaming data,
we instead adapted an online streaming k-means algorithm, ``online Hartigan k-means'' \cite[Appendix B]{McFee:2012}.
This method takes one data point at a time, and applies a weighted update to a selected centroid dependent on the amount of updates that the centroid has received so far.
We adapted the method of \cite[Appendix B]{McFee:2012} for the case of spherical k-means.
K-means is a local optimisation algorithm rather than global, and may be sensitive to the order of presentation of data.
Therefore in order to minimise the effect of order of presentation for the experiments conducted here,
we did not perform the learning in true single-pass streaming mode.
Instead, we performed learning in two passes:
a first streamed pass in which data points were randomly subsampled (using reservoir sampling) and then shuffled before applying PCA whitening and starting the k-means procedure,
and then a second streamed pass in which k-means was further trained by exposing it to all data points.
Our Python code implementation of online streaming spherical k-means is available as supplementary information.

As a further extension to the method, we also tested a \textit{two-layer} version of our feature-learning method, intended to reflect detail across multiple temporal scales.
In this variant, we applied spherical k-means feature learning to a dataset,
and then projected the dataset into that learnt space.
We then downsampled this projected data by a factor of 8 on the temporal scale (by max-pooling, i.e.\ taking the max across each series of 8 frames),
and applied spherical k-means a second time.
The downsampling operation means that the second layer has the potential to learn regularities that emerge across a slightly longer temporal scale.
The two-layer process overall has analogies to deep learning techniques,
most often considered in the context of artificial neural networks \citep{Erhan:2010,Bengio:2013},
and to the progressive abstraction believed to occur towards the higher stages of auditory neural pathways.

%%%%%%%%%%%%%%%%%%%%%%%%%%%%%%%%%%%%%%%%%%%%%%%%
\subsection{Classification and evaluation}

Our full classification workflow started by converting each audio file to a standard sample-rate of 44.1 kHz.
We then calculated Mel spectrograms for each file, using a frame size of 1024 frames with Hamming windowing and no overlap.
We filtered out spectral energy below 500 Hz, a common choice to reduce the amount of environmental noise present,
and then normalised the root-mean-square (RMS) energy in each spectrogram.

For each spectrogram we then optionally applied the noise-reduction procedure that we had found to be useful in our NIPS4B contest submission \citep{Stowell:2013n},
a simple and common median-based thresholding.
This consists of finding the median value for each spectral band in a spectrogram,
then subtracting this median spectrum from every frame,
and setting any resulting negative values to zero.
This therefore preserves only the spectral energy that rises above the median bandwise energy.
In principle it is a good way to reduce the stationary noise background,
but is not designed to cope well with fluctuating noise.
However its simplicity makes it easy to apply across large datasets efficiently.

The Mel spectrograms, either noise-reduced or otherwise, could be used directly as features.
We also tested their reduction to MFCCs (including delta features, making 26-dimensional data),
and their projection onto learned features,
using the spherical k-means method described above.
For the latter option, we tested projections based on single frame as well as on sequences of
2, 3, 4 and 8 frames,
to explore the benefit of modelling short-term temporal variation.
We also tested the two-layer version based on the repeated application to 4-frame sequences across two timescales.

The feature representations thus derived were all time series.
In order to reduce them to summary features for use in the classifier,
we tested two common and simple techniques:
summarising each feature dimension independently by its mean and standard deviation,
or alternatively by its maximum.
These are widespread but are not designed to reflect any temporal structure in the features
(beyond the fine-scale temporal information that is captured by some of our features).
Therefore, for the Mel and MFCC features we also tested summarising by modulation coefficients:
we took the short-time Fourier transform (STFT) along the time axis of our features,
and then downsampled the spectrum to a size of 10 to give a compact representation of the temporal evolution of the features (cf.\ \cite{Lee:2008}).
The multi-frame feature representations already intrinstically included short-term summarisation of temporal variation,
so to limit the overall size of the experiment, for the learned feature representations we only applied the mean-and-standard-deviation summarisation.
Overall we tested six types of non-learned representation against six types of learned representation (Table \ref{tbl:featurecombis}).

To perform classification on our temporally-pooled feature data,
then, we used a random forest classifier
\citep{Breiman:2001}%   title={Random forests}
. Random forests and other tree-ensemble classifiers perform very strongly in a wide range of empirical evaluations
\citep{Caruana:2006}%   title={An empirical comparison of supervised learning algorithms}  ---- finds tree ensembles such as RF good
, and were used by many of the strongest-performing entries to the SABIOD evaluation contests
\citep{Glotin:2013,Fodor:2013,Potamitis:2014}.
For this experiment we used the %\texttt{RandomForestClassifier}
 implementation from the Python \texttt{scikit-learn} project \citep{Pedregosa:2011}.
Note that \texttt{scikit-learn} v0.14 was found to have a specific issue preventing training on large data,
so we used a pre-release v0.15 after verifying that it led to the same results with our smaller datasets.
We did not manually tune any parameters of the classifier.
However, since our experiment covered both single-label and multilabel classification,
we did test three different ways of using the classifier to make decisions:
\begin{enumerate}
	\item	Single-label classification: this assumes that there is only one species present in a recording. It therefore cannot be applied to multilabel datasets, but for single-label datasets it may benefit from being well-matched to the task.
	\item	Binary relevance: this divides the multilabel classification task into many single-label tasks, training one separate classifier for each of the potential output labels \citep{Tsoumakas:2010}. This strategy ignores potential correlations between label occurrence, but potentially allows a difficult task to be approximated as the combination of more manageable tasks. Binary relevance is used e.g.\ by \cite{Fodor:2013}.
	\item	Full multilabel classification: in this approach, a single classifier (here, a single random forest) is trained to make predictions for the full multi-label situation. Predicting presence/absence of every label simultaneously can be computationally difficult compared against a single-label task, and may require larger training data volumes, but represents the full situation in one model \citep{Tsoumakas:2010}.
\end{enumerate}
For each of these methods the outputs from the classifier are per-species probabilities.
We tested all of our datasets using the full multilabel classifier,
then for comparison we tested the single-label datasets using the single-label classifier,
and the multi-label datasets using the binary-relevance classifier.

\begin{figure}[tp]
	\begin{center}
	\includegraphics [width=0.48\textwidth,clip, trim=0mm 0mm 0mm 0mm]{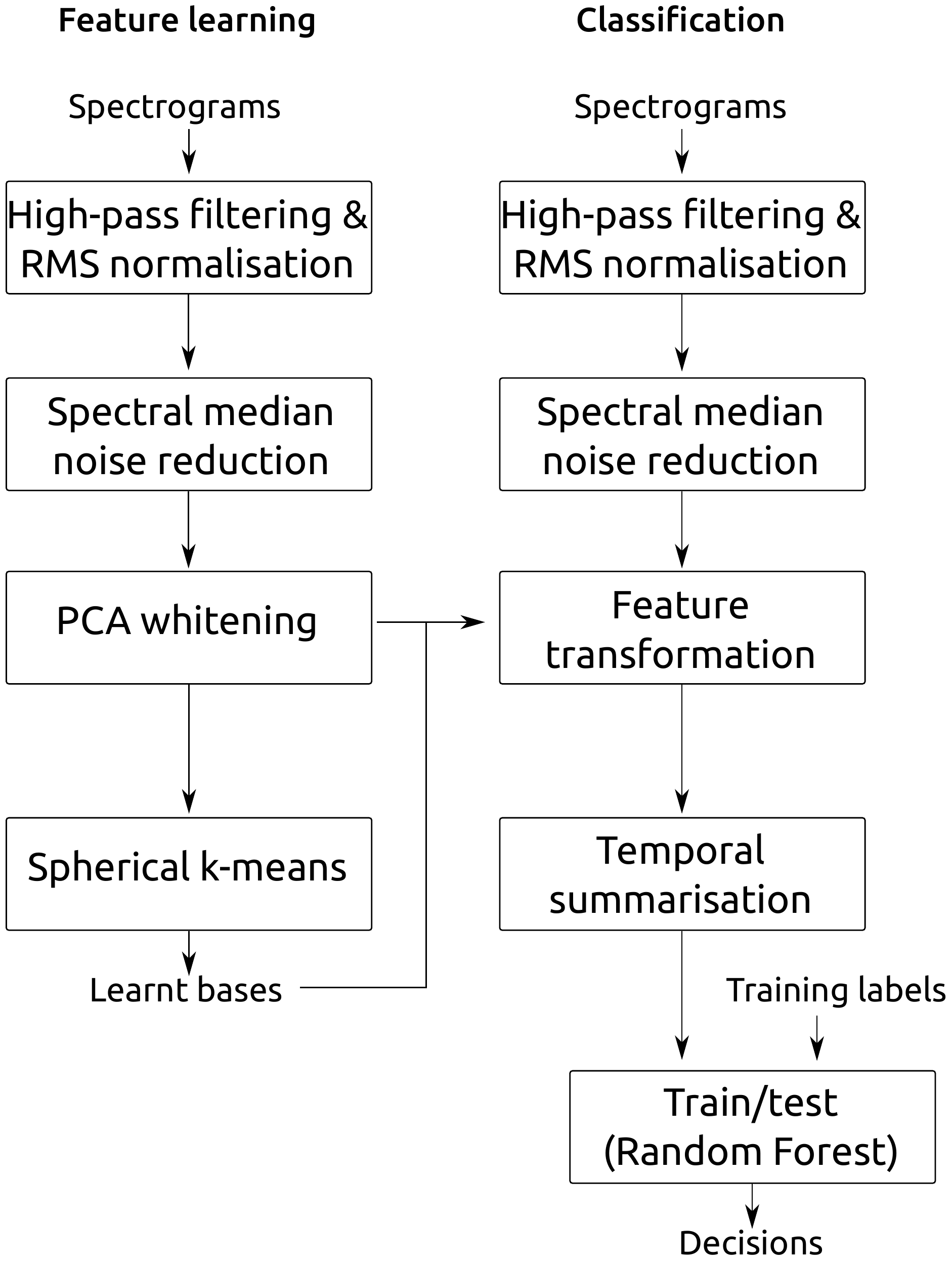}
	\end{center}
	\caption{Summary of the classification workflow, here showing the case where single-layer feature learning is used.}
	\label{fig:kflworkflow}
\end{figure}

Some of our datasets contain long audio recordings, yet none of the annotations indicate which point(s) in time each species is heard.
This is a common format for annotations: for example, the \textit{bldawn} annotations are derived directly from the archival metadata,
which was not designed specifically for automatic classification.
Long audio files present an opportunity to make decisions either for the entire file as one scene,
or in smaller ``decision windows'', for which the decisions are then pooled to yield overall decisions.
We tested this empirically, using decision windows of length 1, 5 or 60 seconds or the whole audio.
Each decision window was treated as a separate datum for the purposes of training and testing the classifier,
and then the decisions were aggregated per audio file using either mean or maximum.
The mean probability of a species across all the decision windows is a reasonable default combination;
we compared this against the maximum with the motivation that if a bird is heard only at one point in the audio,
and this leads to a strong detection in one particular decision window, then such a strong detection should be the overriding factor in the overall decision.
For some datasets (\textit{nips4b}) we did not test long windows since all audio files were short;
while for \textit{lifeclef2014} we used only whole-audio classification because of the runtime costs of evaluating these combinations over this largest dataset.

We performed feature learning, training and testing separately for each of our four datasets,
using the appropriate two- or threefold crossvalidation described above,
and across all combinations of the feature settings we have just described.
%We also performed these stages separately for the cross-condition and data-augmentation versions of the \textit{bldawn} train/test scheme.
 Fig. \ref{fig:kflworkflow} summarises the main stages of the workflow described.

We evaluated the performance in each experimental run using two measures, the Area Under the ROC Curve (AUC) and the Mean Average Precision (MAP).
The AUC statistic is an evaluation measure for classification/detection systems which has many desirable properties \citep{Fawcett:2006}:
unlike raw accuracy, it is not affected by ``unbalanced'' datasets having an uneven mixture of true-positive and true-negative examples;
and it has a standard probabilistic interpretation,
in that the AUC statistic tells us the probability that the algorithm will rank a random positive instance
higher than a random negative instance.
Chance performance is always 50\% for the AUC statistic.
The AUC is a good statistic to use to evaluate the output from a probabilistic classifier in general,
but for user-facing applications, which may for example show a ranked list of possible hits,
the Mean Average Precision (MAP) statistic leads to an evaluation which relates more closely to user satisfaction with the ranked list.
The key difference in effect is that the AUC statistic applies an equal penalty for a mis-ordering at any position on the ranked list,
whereas the MAP statistic assigns greater penalties higher up the ranking \citep{Yue:2007}.
We therefore calculate both evaluation statistics.

\begin{table}[t]
\begin{center}
\resizebox{\textwidth}{!}{
\begin{tabular}{ l | l l r }
	Label	&	Features	&	Summarisation	&	Dimension \\
	\hline
\texttt{mfcc-ms}                &	MFCCs (+deltas)	&	Mean \& stdev	&	52 \\
\texttt{mfcc-maxp}              &	MFCCs (+deltas)	&	Max         &	26 \\
\texttt{mfcc-modul}             &	MFCCs (+deltas)	&	Modulation coeffs	&	260 \\
\texttt{melspec-ms}             &	Mel spectra    	&	Mean \& stdev	&	80 \\
\texttt{melspec-maxp}           &	Mel spectra    	&	Max	&	40 \\
\texttt{melspec-modul}          &	Mel spectra    	&	Modulation coeffs	&	400 \\
\texttt{melspec-kfl1-ms}        &	Learned features, 1 frame 	&	Mean \& stdev	&	1000 \\
\texttt{melspec-kfl2-ms}        &	Learned features, 2 frames	&	Mean \& stdev	&	1000 \\
\texttt{melspec-kfl3-ms}        &	Learned features, 3 frames	&	Mean \& stdev	&	1000 \\
\texttt{melspec-kfl4-ms}        &	Learned features, 4 frames	&	Mean \& stdev	&	1000 \\
\texttt{melspec-kfl8-ms}        &	Learned features, 8 frames	&	Mean \& stdev	&	1000 \\
\texttt{melspec-kfl4pl8kfl4-ms}	&	Learned features, 4 frames, two-layer 	&	Mean \& stdev	&	1000 \\

\end{tabular}
}
\end{center}
\caption{The twelve combinations of feature-type and feature-summarisation tested. The feature-type and feature-summarisation method jointly determine the dimensionality of the data input to the classifier.}
\label{tbl:featurecombis}
\end{table}

To test for significant differences in the performance statistics, we applied a general linear model (GLM)
using the \texttt{lme4} package \citep{Bates:2014} for R 2.15.2 \citep{R}.
We focused primarily on the AUC for significance testing,
since the AUC and MAP statistics are related analyses of the same data.
Since AUC is bounded in the range [0,1], we applied the GLM in the logistic domain:
note that given the probabilistic interpretation of AUC,
the logistic model is equivalent to applying a linear GLM to odds ratios,
a fact which facilitates interpretation.
Every experimental run for a given dataset used the same set of folds, so we used a repeated-measures version of the GLM with the ``fold index'' as the grouping variable.
We tested for individual and pairwise interactions of our five independent categorical variables, which were as follows:
\begin{itemize}
	\item	choice of feature set and temporal summarisation method, testing the 12 configurations listed in Table \ref{tbl:featurecombis}:
	\item	noise reduction on vs.\ off;
	\item	classifier mode (multilabel vs.\ either single-label or binary-relevance);
	\item	decision pooling window duration (1, 5, or 60 seconds or whole audio);
	\item	decision pooling max vs.\ mean.
\end{itemize}
Combinatorial testing of all these configurations resulted in $12 \times 2 \times 2 \times 4 \times 2 = 384$ crossvalidated classification experiments for the \texttt{bldawn} and \texttt{xccoverbl} datasets.
For the other datasets we did not test all four pooling durations, for reasons given above:
the number of crossvalidated experiments was thus $192$ for \texttt{nips4b} and $96$ for \texttt{lifeclef2014}.
Since the tests of \texttt{lifeclef2014} did not vary decision pooling, decision pooling factors were not included in that GLM.
We considered effects to be significant when the 95\% confidence interval calculated from the GLM excluded zero,
in which cases we report the estimated effects as differences in odds-ratios.

\subsection{Additional tests}

The \textit{bldawn} dataset has relatively few annotations, since it only consists of 60 items.
We therefore wanted to explore the use of auxiliary information from other sources to help improve recognition quality,
in particular using the \textit{xccoverbl} dataset, which has strong overlap in the list of species considered.
In further tests we tested three ways of using this additional data:
\begin{enumerate}
	\item	\textbf{Cross-condition training}, meaning training on one dataset and testing on the other.
The two datasets have systematic differences---for example, \textit{xccoverbl} items are annotated with only one species each, and are generally shorter---and so we did not expect this to yield very strong results.
	\item	\textbf{Data augmentation} for the feature learning step, meaning that feature learning is conducted using the training data for the \textit{bldawn} as well as all of the \textit{xccoverbl} data. This gives a larger and more varied pool of data for the feature learning step, which we expected to give a slight improvement to the results of feature learning.
	\item	\textbf{Data augmentation} for feature learning and also for training. Although the systematic differences mean the \textit{xccoverbl} training data might not guide the classifier in the correct way,
it holds many more species annotations for the species of interest, in a wider set of conditions.
We expected the combined training would provide stronger generalisation performance.
\end{enumerate}
We evaluated these train/test conditions as separate evaluation runs.

We also wanted to distinguish between two possible explanations for any difference between the performance of the different feature sets:
was it due to intrinsic characteristics of the features, or more simply due to the dramatic differences in feature dimensionality (which ranged 26 to 1000; see Table \ref{tbl:featurecombis})?
Differences in dimensionality might potentially give more degrees of freedom to the classifier without necessarily capturing information in a useful way.
In order to test this, we ran a version of our test in which for each experimental run we created a random projection matrix which projected the feature set to a fixed dimensionality of 200.
For MFCC/Mel features this was a simple form of data expansion,
while for learned features it was a form of data reduction.
By standardising the feature dimensionality,
this procedure decoupled the nature of the feature set from the degrees of freedom available to the classifier.
We ran this test using the \textit{nips4b} dataset.

%%%%%%%%%%%%%%%%%%%%%%%%%%%%%%%%%%%%%%%%%%%%%%%
\section{Results}
\label{sec:results}

\begin{figure}[tp]
	\begin{center}
	\includegraphics [width=0.49\textwidth,clip, trim=2.5mm 2.5mm 2.5mm 2.5mm]{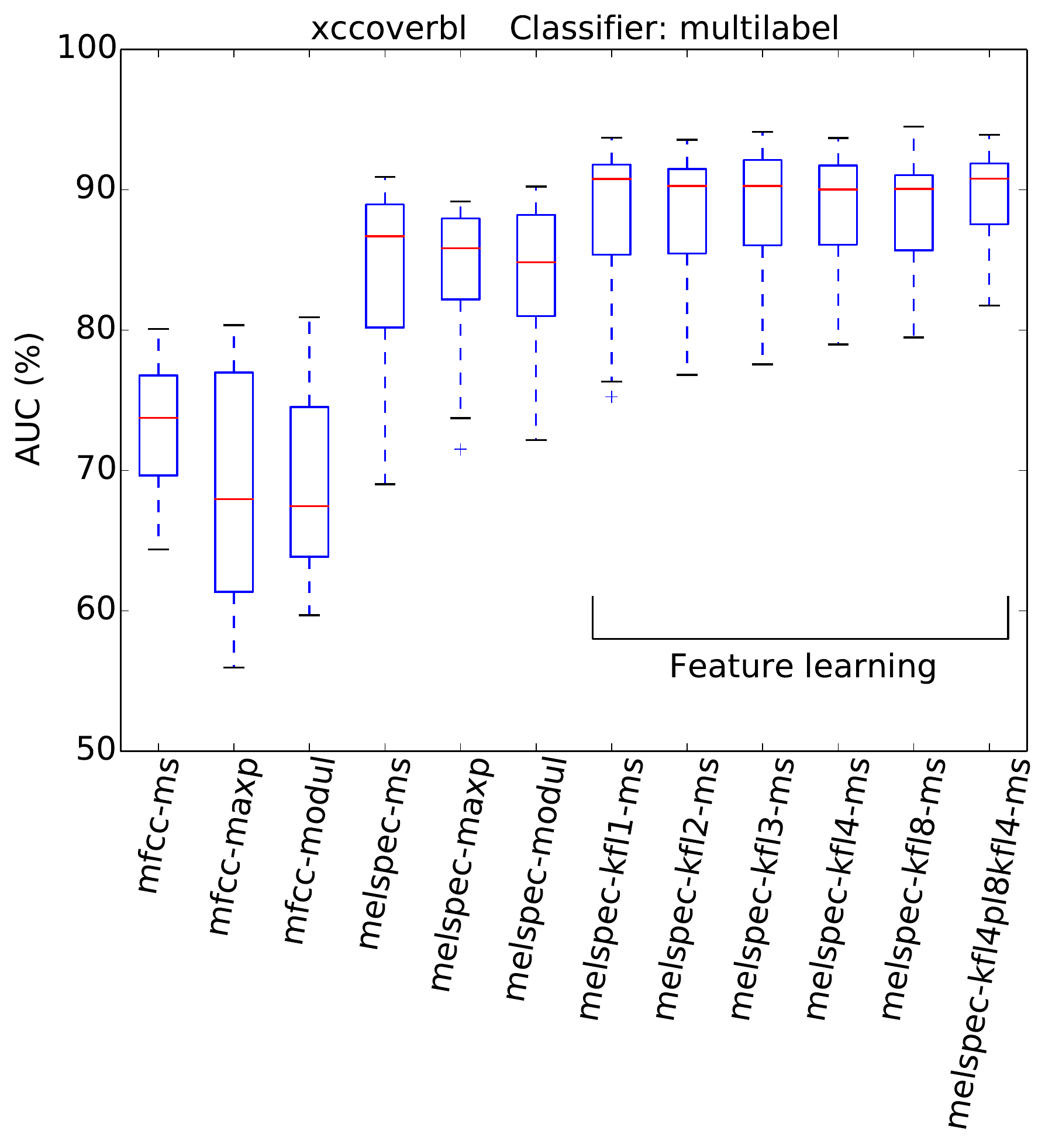}
	\includegraphics [width=0.49\textwidth,clip, trim=2.5mm 2.5mm 2.5mm 2.5mm]{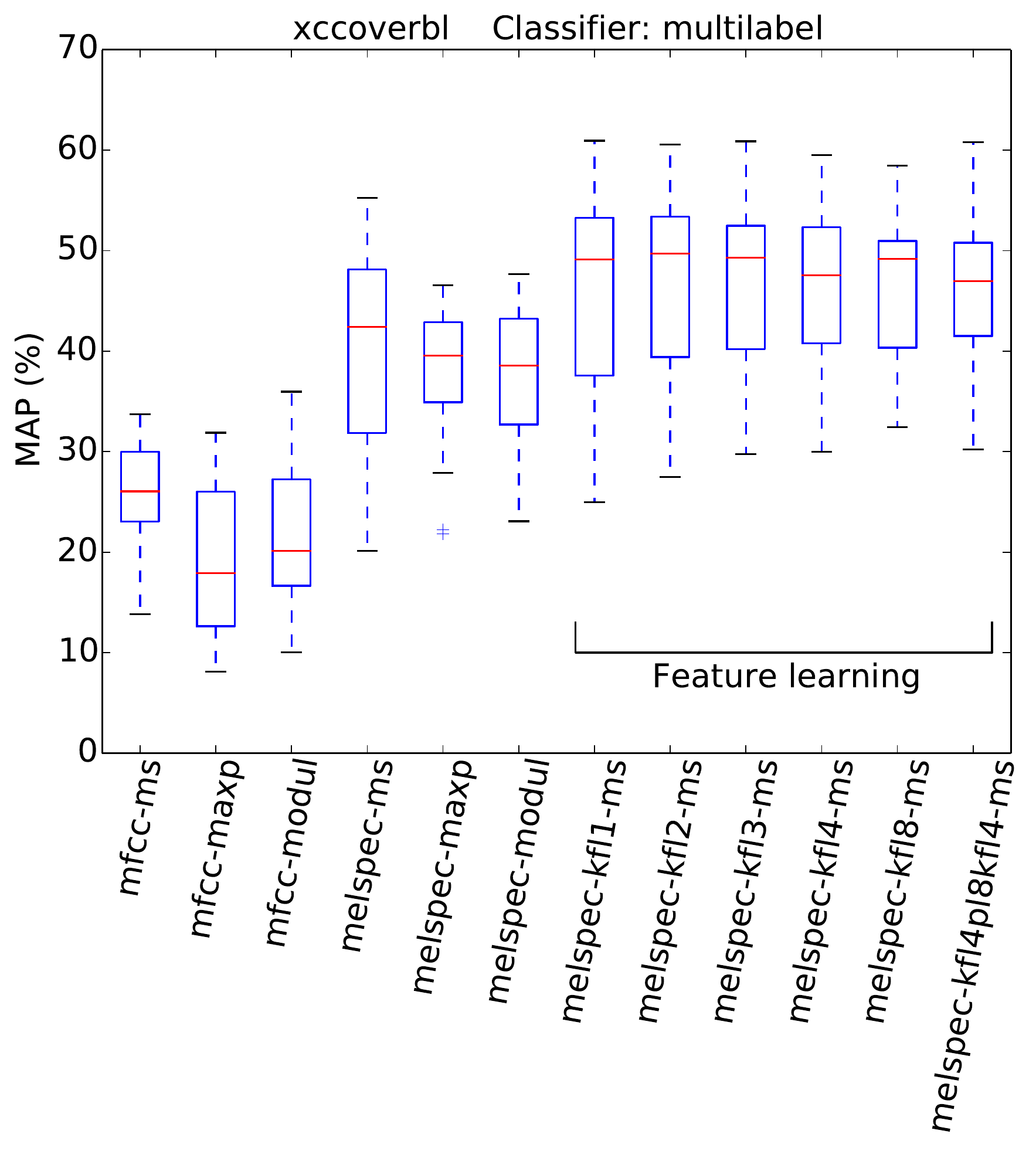}
	\\
	\includegraphics [width=0.49\textwidth,clip, trim=2.5mm 2.5mm 2.5mm 2.5mm]{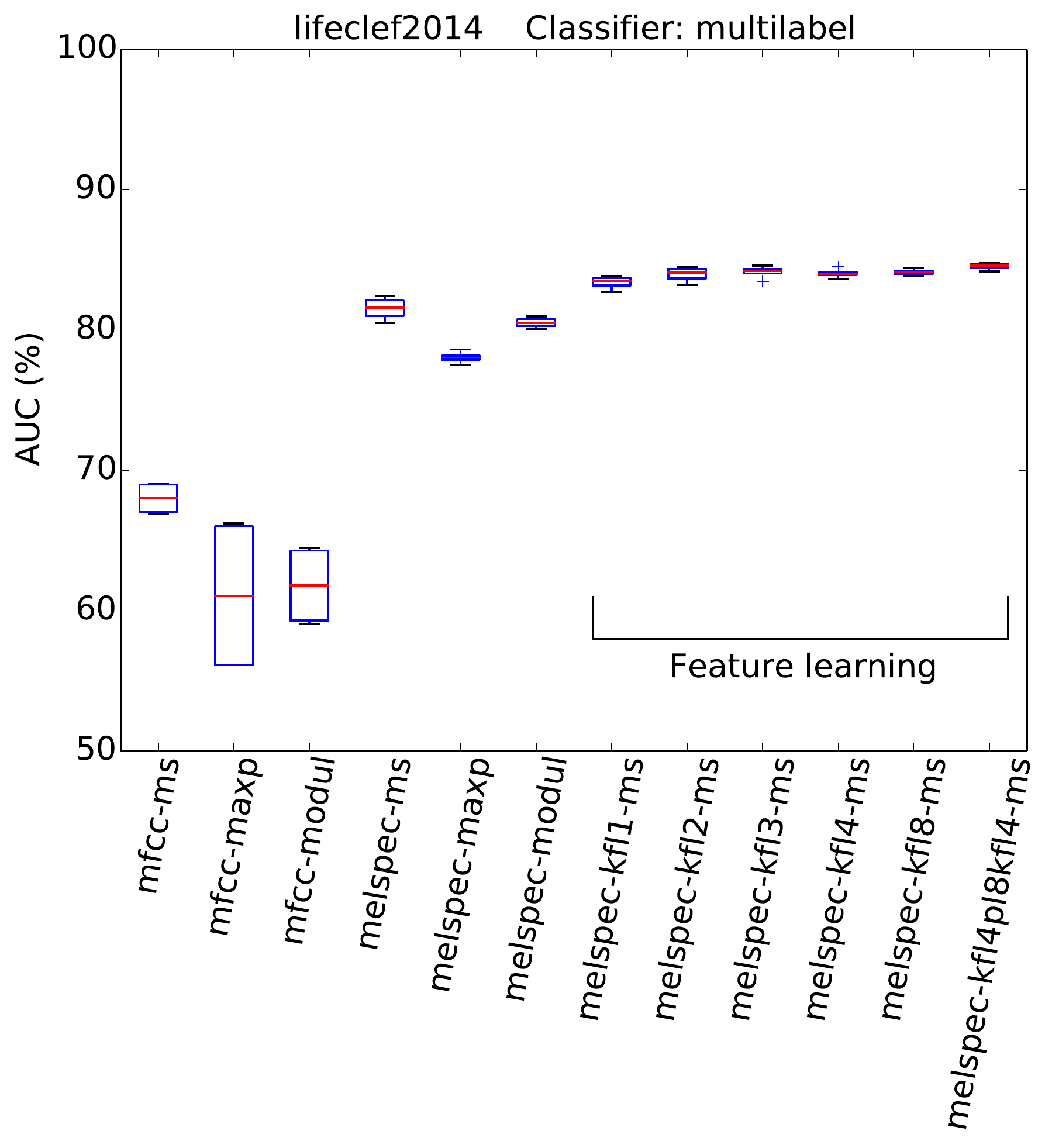}
	\includegraphics [width=0.49\textwidth,clip, trim=2.5mm 2.5mm 2.5mm 2.5mm]{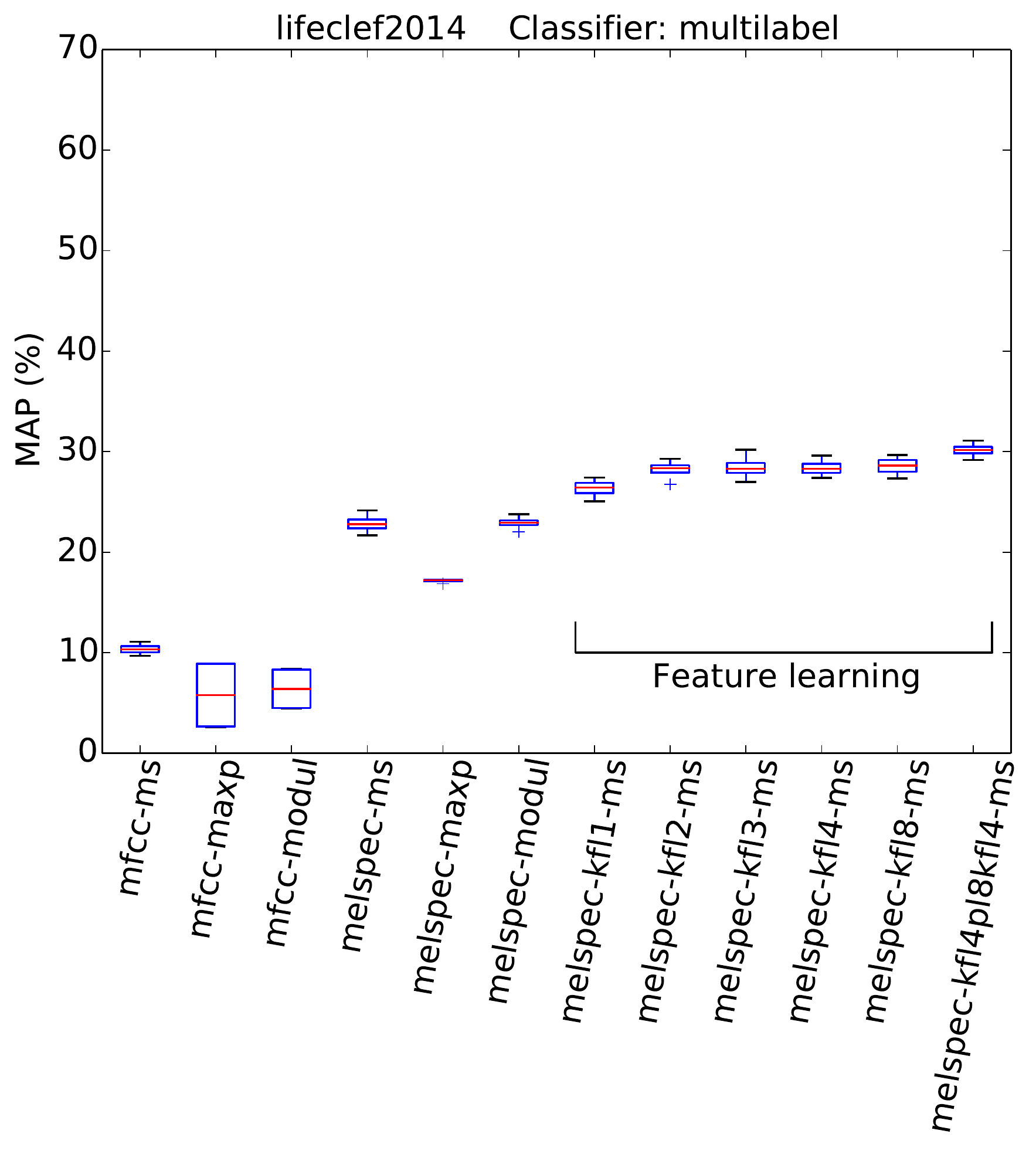}
	\end{center}
	\caption{AUC and MAP statistics, summarised for each feature-type tested---here for the two single-label datasets, %
using the full multilabel classifier. %
Each column in the boxplot summarises the crossvalidated scores attained over many combinations of the other configuration settings tested %
(for the full multi-class classifier only). %
The ranges indicated by the boxes therefore do not represent random variation due to training data subset, but systematic variation due to classifier configuration. %
Figure \ref{fig:resultsfeatm} plots the same for the multilabel datasets. %
}
	\label{fig:resultsfeats}
\end{figure}

\begin{figure}[tp]
	\begin{center}
	\includegraphics [width=0.49\textwidth,clip, trim=2.5mm 2.5mm 2.5mm 2.5mm]{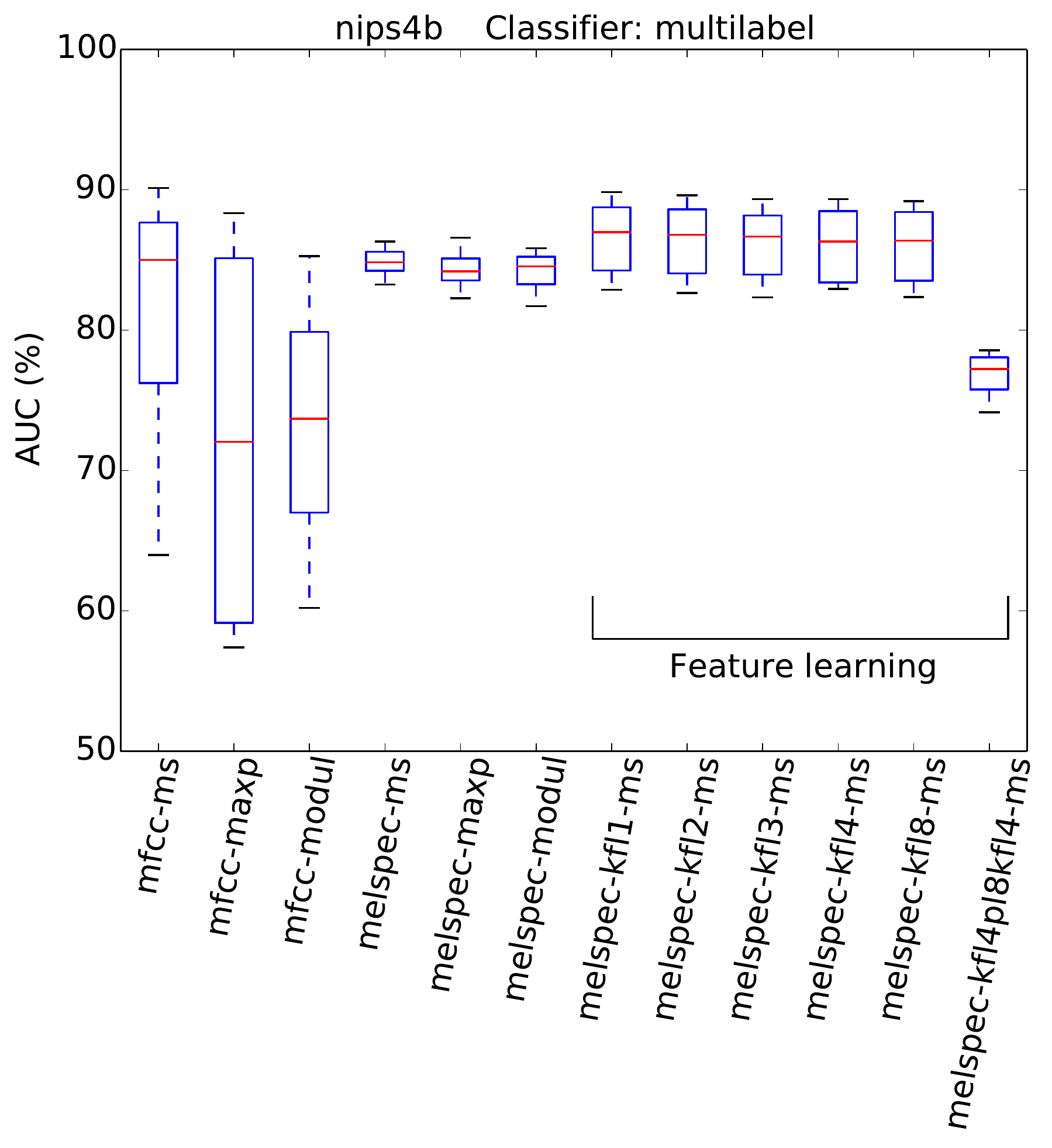}
	\includegraphics [width=0.49\textwidth,clip, trim=2.5mm 2.5mm 2.5mm 2.5mm]{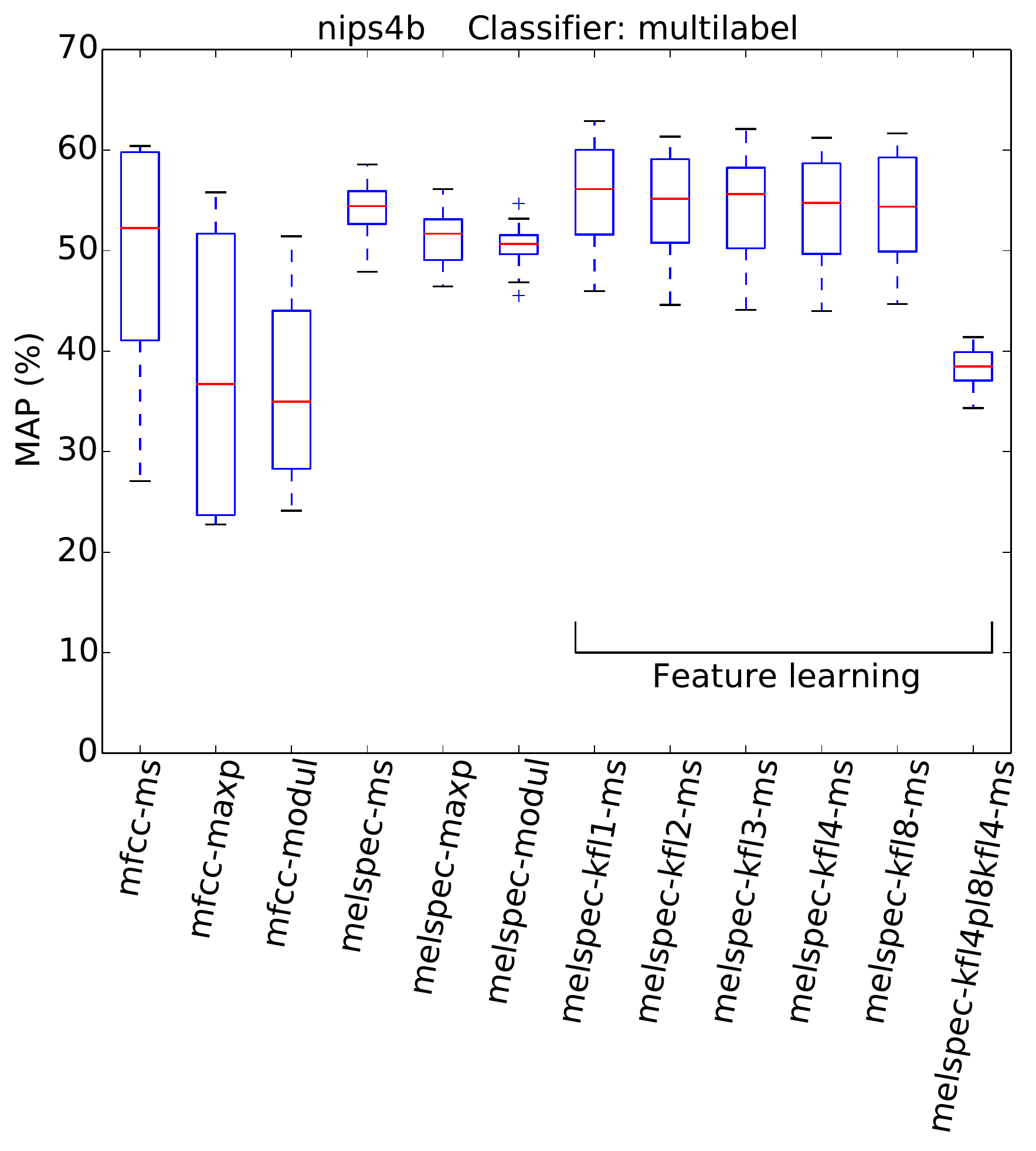}
	\\
	\includegraphics [width=0.49\textwidth,clip, trim=2.5mm 2.5mm 2.5mm 2.5mm]{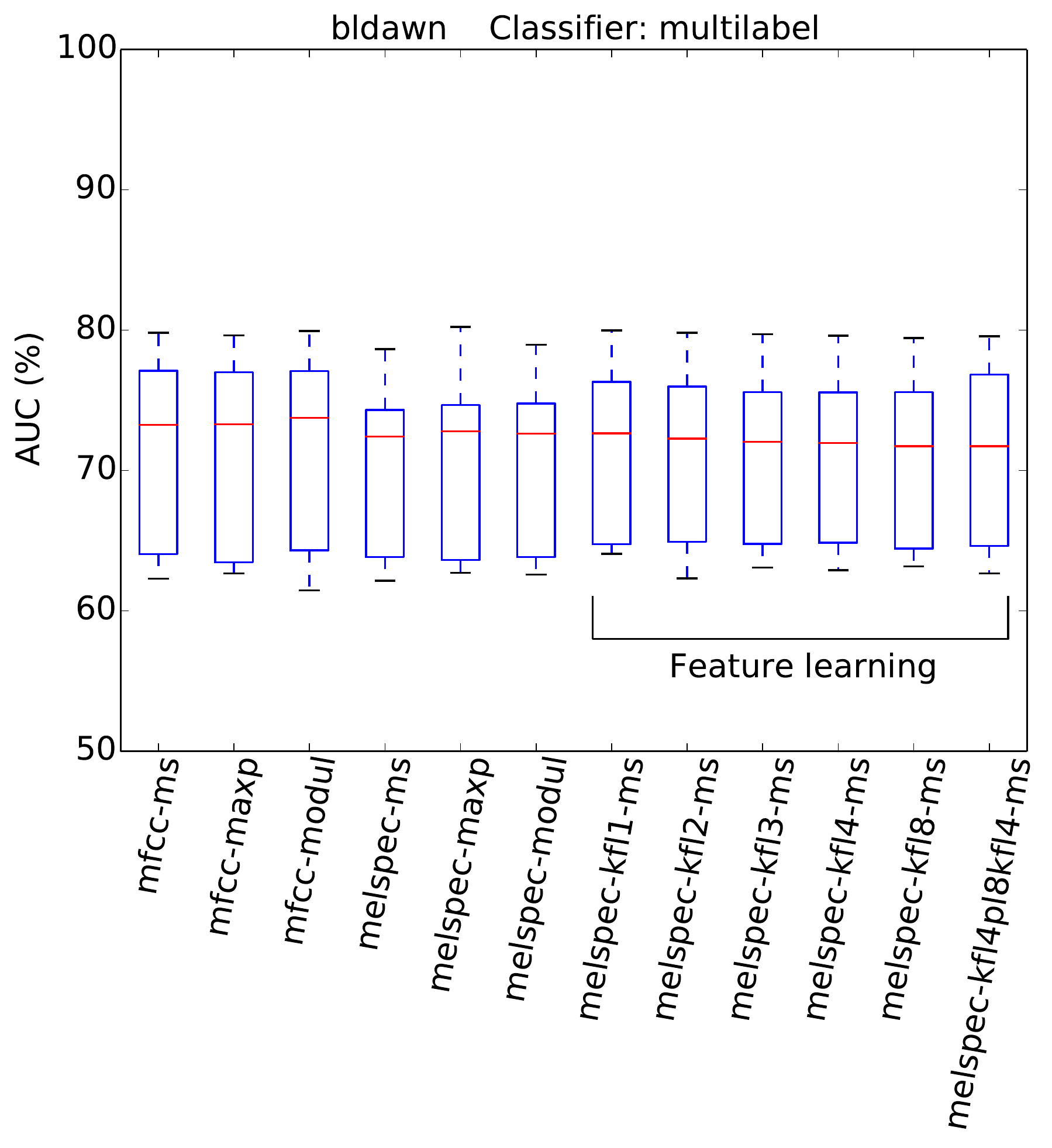}
	\includegraphics [width=0.49\textwidth,clip, trim=2.5mm 2.5mm 2.5mm 2.5mm]{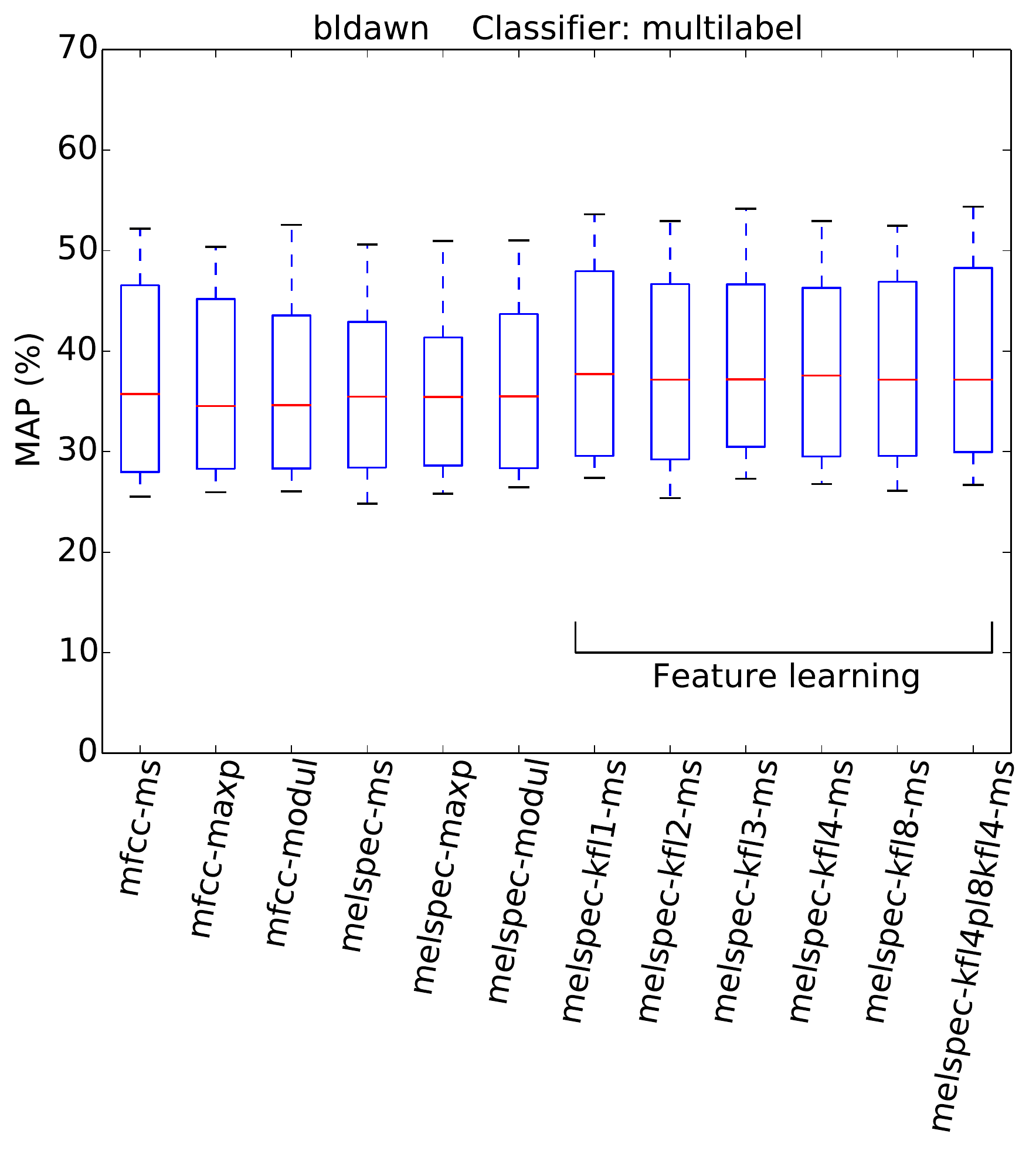}
	\end{center}
	\caption{As Fig. \ref{fig:resultsfeats}, but for the two multilabel datasets, with the multilabel classifier.}
	\label{fig:resultsfeatm}
\end{figure}

\begin{figure}[tp]
	\begin{center}
	\includegraphics [width=0.49\textwidth,clip, trim=2.5mm 2.5mm 2.5mm 2.5mm]{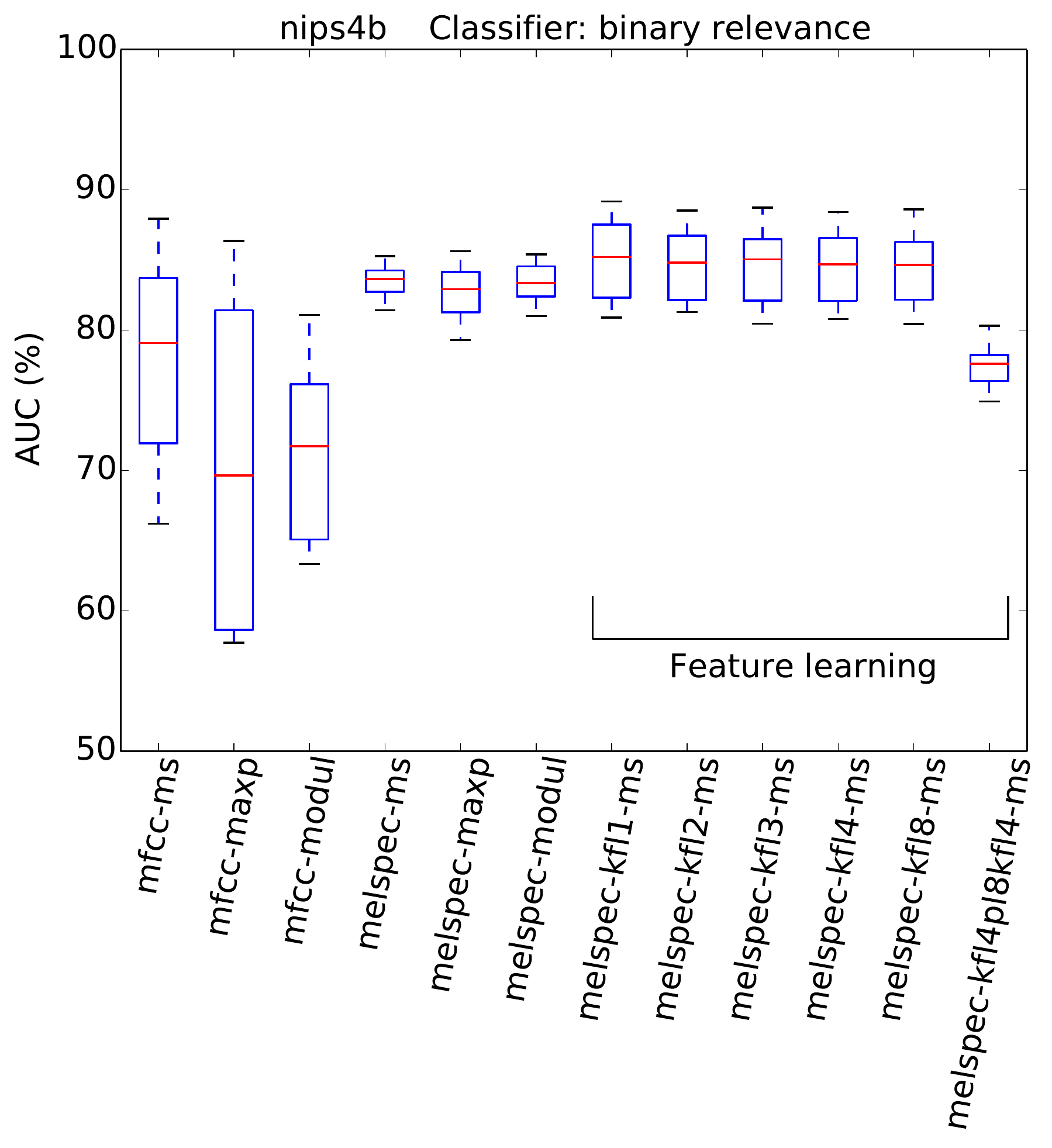}
	\includegraphics [width=0.49\textwidth,clip, trim=2.5mm 2.5mm 2.5mm 2.5mm]{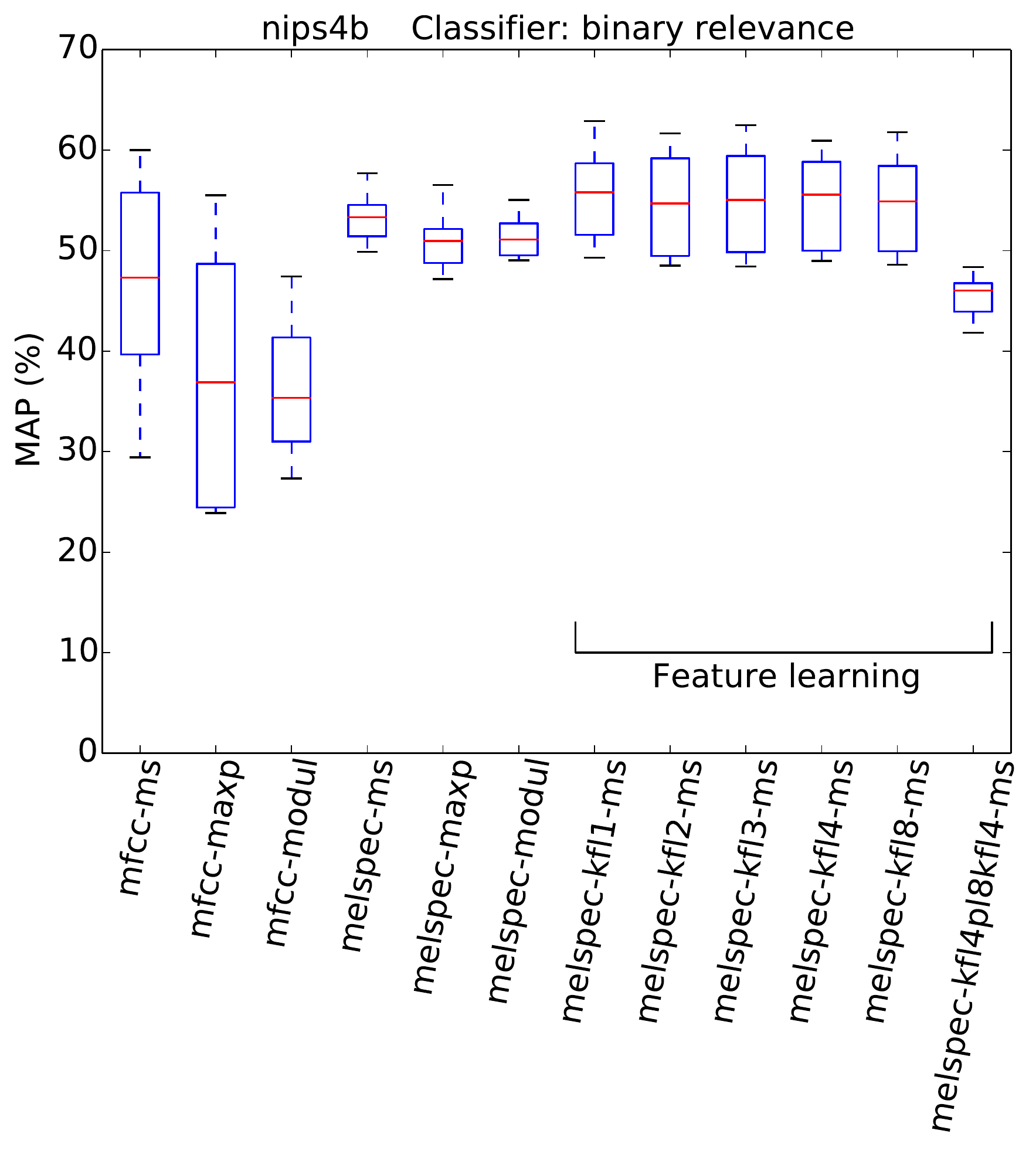}
	\\
	\includegraphics [width=0.49\textwidth,clip, trim=2.5mm 2.5mm 2.5mm 2.5mm]{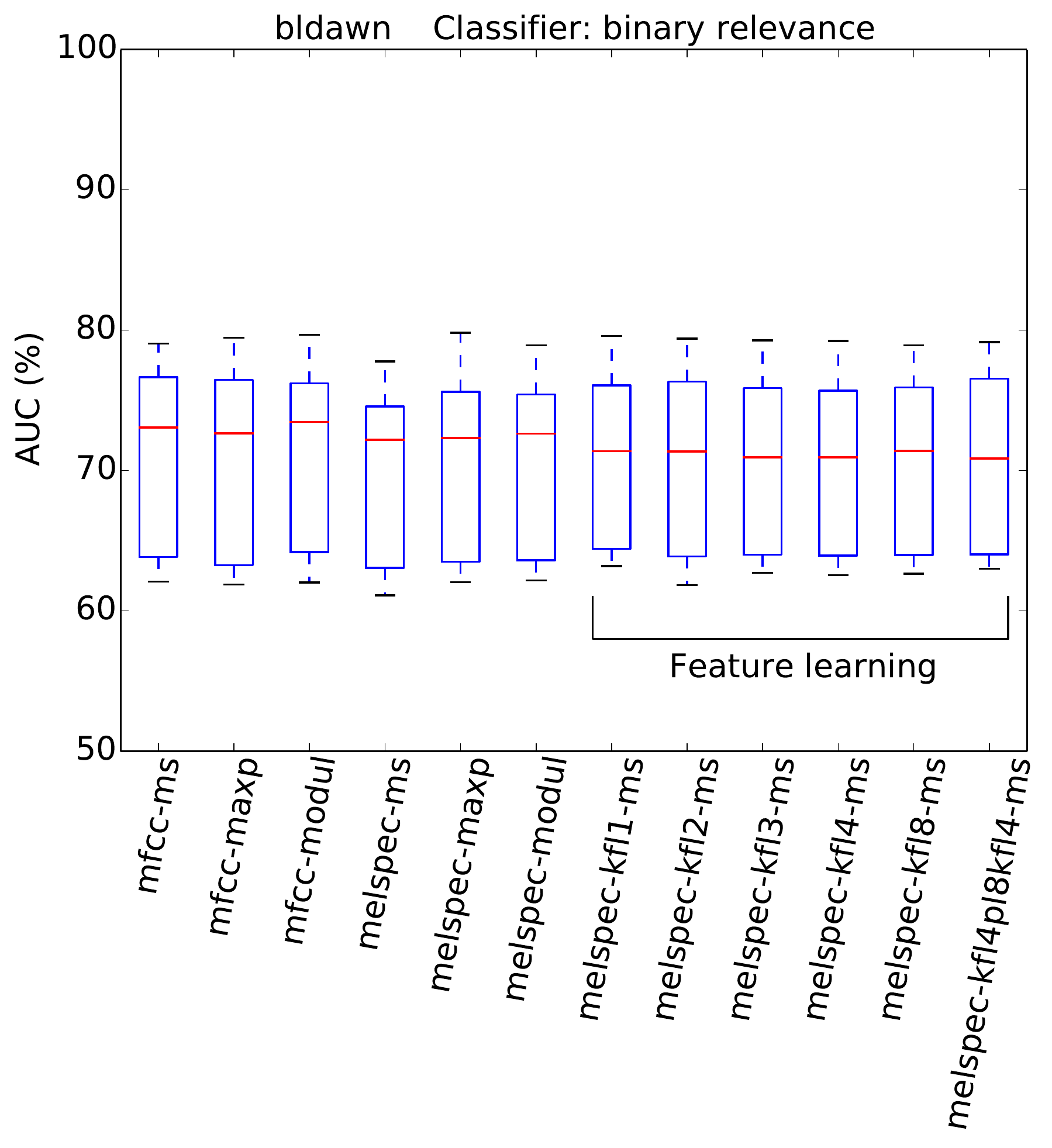}
	\includegraphics [width=0.49\textwidth,clip, trim=2.5mm 2.5mm 2.5mm 2.5mm]{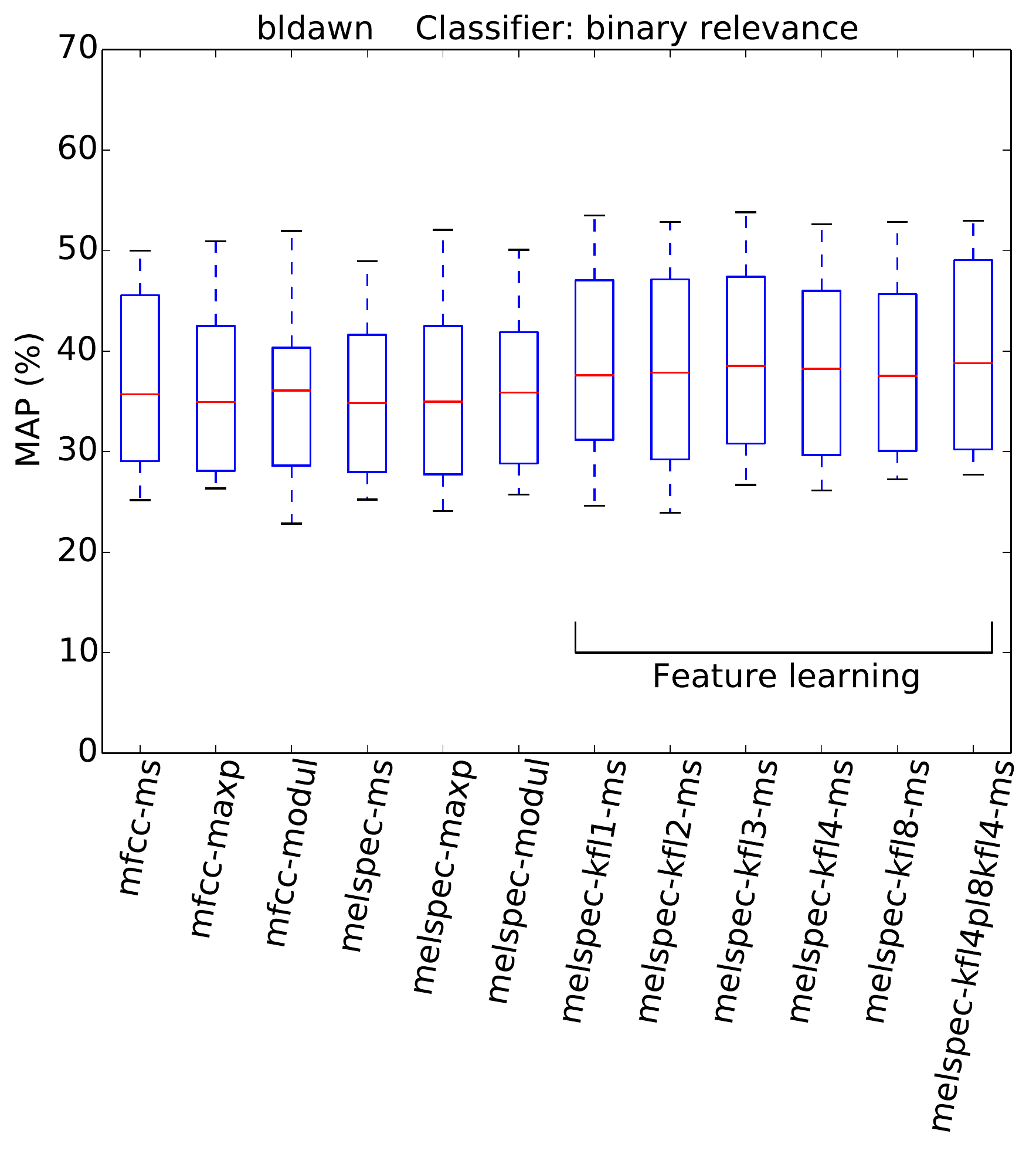}
	\end{center}
	\caption{As Fig. \ref{fig:resultsfeats}, but for the two multilabel datasets, with the binary relevance classifier.}
	\label{fig:resultsfeatrfeach}
\end{figure}

\begin{table}[t]
\begin{center}
\resizebox{\textwidth}{!}{
\begin{tabular}{ l l | r r r r }
	Factor & Factor value  & \texttt{nips4b} & \texttt{xccoverbl} & \texttt{bldawn} & \texttt{lifeclef2014}    \\
	\hline

\hline
featureset        & {mfcc-maxp}               & \textbf{* -0.59} & \textbf{* -0.29} & -0.03 & \textbf{* -0.30} \\
(vs.\ {mfcc-ms})        & {mfcc-modul}               & \textbf{* -0.55} & \textbf{* -0.45} & \textbf{* -0.12} & \textbf{* -0.27} \\
        & {melspec-ms}               & \textbf{* 0.10} & \textbf{* 1.01} & \textbf{* -0.04} & \textbf{* 0.73} \\
        & {melspec-maxp}               & -0.01 & \textbf{* 0.82} & -0.03 & \textbf{* 0.52} \\
        & {melspec-modul}               & 0.02 & \textbf{* 0.82} & -0.03 & \textbf{* 0.67} \\
        & {melspec-kfl1-ms}               & \textbf{* 0.26} & \textbf{* 1.43} & 0.01 & \textbf{* 0.86} \\
        & {melspec-kfl2-ms}               & \textbf{* 0.25} & \textbf{* 1.36} & -0.02 & \textbf{* 0.90} \\
        & {melspec-kfl3-ms}               & \textbf{* 0.23} & \textbf{* 1.44} & -0.00 & \textbf{* 0.92} \\
        & {melspec-kfl4-ms}               & \textbf{* 0.20} & \textbf{* 1.40} & -0.00 & \textbf{* 0.91} \\
        & {melspec-kfl8-ms}               & \textbf{* 0.21} & \textbf{* 1.39} & -0.01 & \textbf{* 0.91} \\
        & {melspec-kfl4pl8kfl4-ms}               & \textbf{* -0.36} & \textbf{* 1.40} & -0.00 & \textbf{* 0.95} \\
\hline
noisered.        & {on}               & \textbf{* -0.64} & \textbf{* -0.20} & -0.01 & \textbf{* 0.06} \\
\hline
pooldur        & {1}               & \textbf{* -0.23} & \textbf{* -0.15} & -0.02 &  \\
(vs.\ none)        & {5}               &  & \textbf{* -0.15} & -0.04 &  \\
        & {60}               &  & 0.04 & -0.00 &  \\
\hline
dpoolmode        & {mean}               & 0.03 & \textbf{* 0.07} & 0.01 &  \\
\hline
classif        & {binary relevance}               & \textbf{* -0.28} &  & \textbf{* -0.05} & \\
(vs.\ multi)        & {single-label}               &  & 0.05 &  & 0.01 \\

\end{tabular}
}
\end{center}
\caption{First-order effect sizes, estimated by our GLM as linear changes to the AUC odds ratio. %
Significant results are marked in bold with an asterisk, %
judged relative to a baseline category indicated in the first column. %
Positive values represent an improvement over the baseline. %
Empty cells indicate combinations that were not tested, as described in the text. %
}
\label{tbl:effectsizes}
\end{table}

Recognition performance was generally strong (Fig. \ref{fig:resultsfeats}, Fig. \ref{fig:resultsfeatm}, Table \ref{tbl:effectsizes}), given the very large number of classes to distinguish (at least 77).
The AUC and MAP performance measures both led to very similar rankings in our experiments.

% MAIN OBSERVATION: KFL > MEL > MFCC
The strongest effect found in our tests was the effect of feature type,
with a broad tendency for MFCCs to be outperformed by Mel spectra, and both of these outperformed by learned features.
For the largest dataset, \textit{lifeclef2014}, feature learning led to classification performance up to 85.4\% AUC,
whereas without feature learning the performance peaked at 82.2\% for raw Mel spectra or 69.3\% for MFCCs.
This pattern was clear for all datasets except \textit{bldawn}.
Compared against the baseline standard configuration \texttt{mfcc-ms}, % (MFCCs summarised by mean and standard deviation),
switching to learned features provided all the strongest observed boosts in recognition performance
(Table \ref{tbl:effectsizes}).
The effect was particularly strong for the two single-label datasets, \textit{xccoverbl} and \textit{lifeclef2014} (effect size estimates $\geq 0.86$ for all feature-learning variants).
For \textit{nips4b} there was a milder effect ($\approx 0.25$),
except for the two-layer version which had a significant negative effect (-0.36).
Conversely, the two-layer version achieved strongest performance on the largest dataset (\textit{lifeclef2014}).
These facts together suggest that the performance impairment for \textit{nips4b} was due to the relatively small size of the dataset,
since deeper models typically require more data \citep{Coates:2012}.
That aside, the performance differences between variants of our feature-learning method were small.
For all datasets except \textit{bldawn},
the switch from MFCCs to raw Mel spectral features also provided a strong boost in performance,
though not to the same extent as did the learned features.
Across those three datasets, mean-and-standard-deviation summarisation consistently gave the strongest performance
over our two alternatives (i.e.\ maximum or modulation coefficients).

% Talk about bldawn
None of the above tendencies are discernible in the results for \textit{bldawn},
for which all methods attain the same performance.
The classifier can reach over 80\% AUC (50\% MAP), which is far above chance performance,
but not as strong as for the other datasets,
nor showing the pattern of differentiation between the configurations.
The relatively low scores reflect a relatively low ability to generalise,
demonstrated by the observation that the trained systems attained very strong scores when tested on their training data ($>99.95$\% in all cases, and 100\% in most).
This outcome is typical of systems trained with an amount of ground-truth annotations which is insufficient to represent the full range of variation within the classes.

% classifmode
The choice of classifier mode showed moderate but consistent effects across all the combinations tested.
For multilabel datasets, a decrease in AUC was observed
by switching to the ``binary relevance'' approach to classification.
Note however that this difference is more pronounced for AUC than for MAP (Fig. \ref{fig:resultsfeatrfeach}).
For single-label datasets, no significant effect was observed,
with a very small boost in AUC
by switching from the multilabel classifier to the single-label classifier. % (Fig. \ref{fig:resultsfeatrfsgl}).

% dpool
Splitting the audio into decision windows and then combining the outcomes generally had a negative or negligible effect on outcomes;
however, using mean (rather than maximum) to aggregate such decisions had a mild positive effect (significant only for \textit{xccoverbl}).
Looking at the second-order interaction did not find any synergistic positive effects of using mean-pooling and a particular window length.
% nr
Activating noise reduction showed an inconsistent effect, significantly impairing performance on \textit{nips4b} (and to a lesser extent \textit{xccoverbl})
while slightly improving performance on \textit{lifeclef2014}.

%\subsection{Additional tests}

\begin{figure}[tp]
	\begin{center}
	\includegraphics [width=0.49\textwidth,clip, trim=2.5mm 67.7mm 2.5mm 2.5mm]{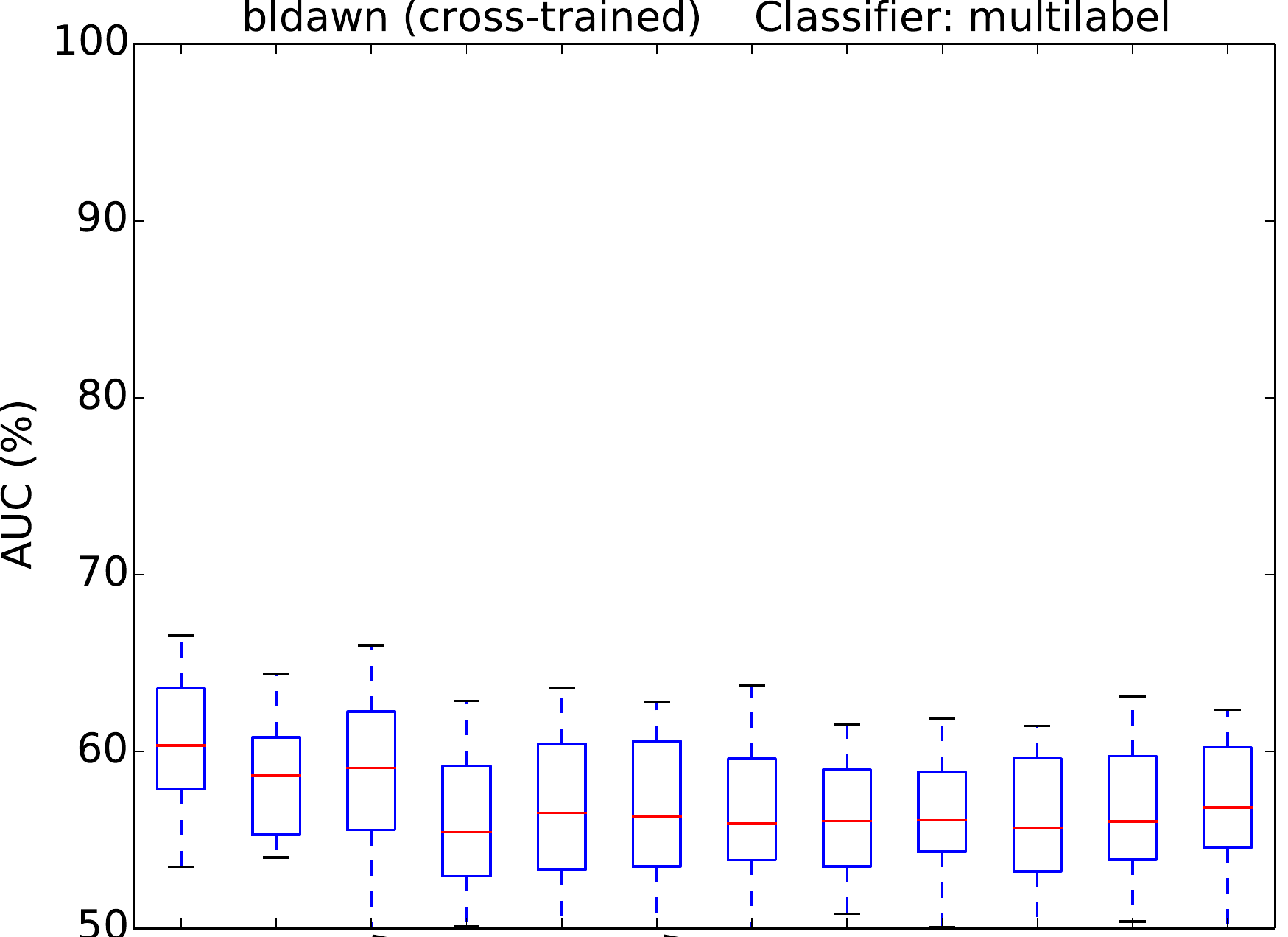}
	\includegraphics [width=0.49\textwidth,clip, trim=2.5mm 67.7mm 2.5mm 2.5mm]{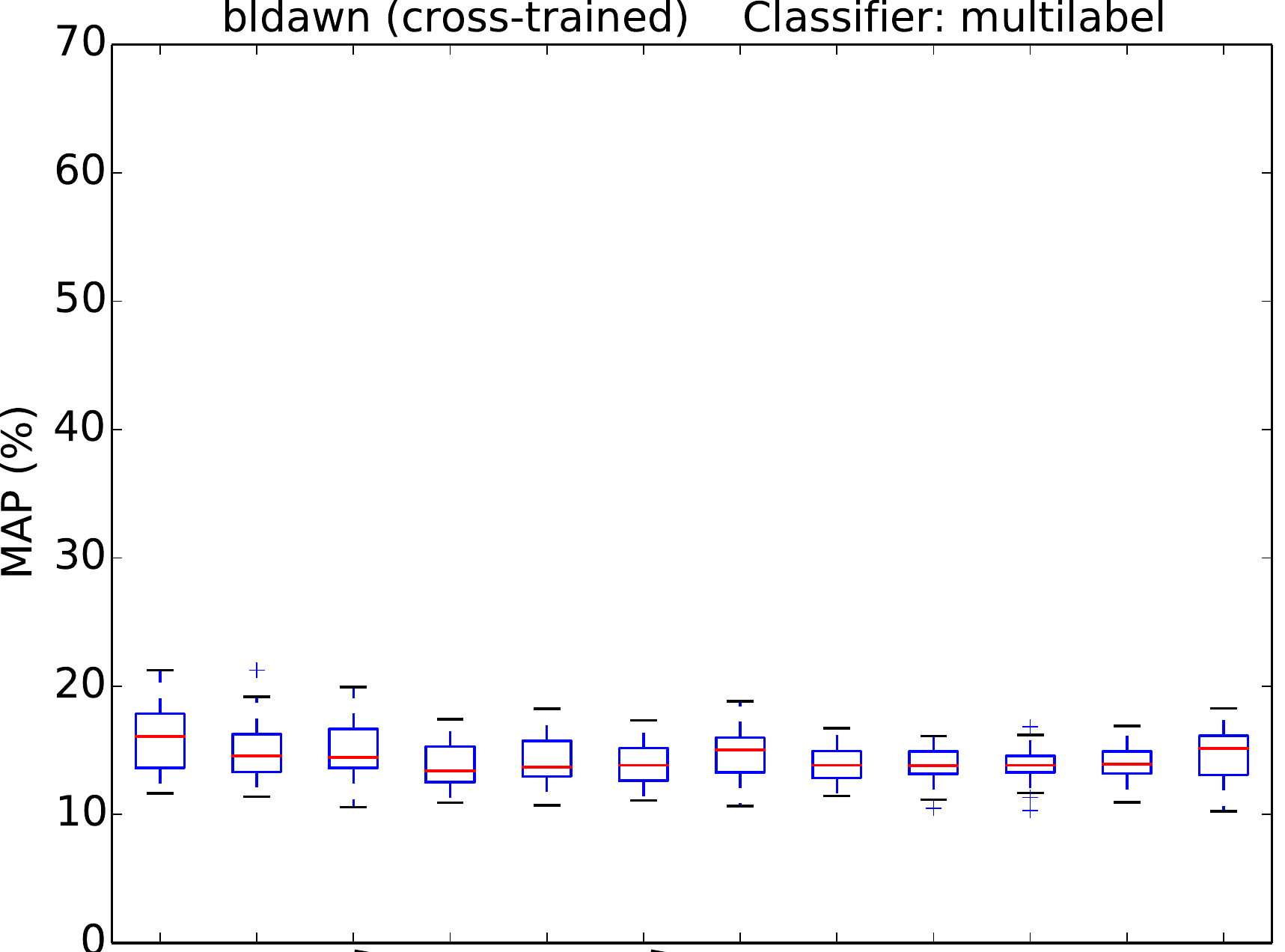}
	\\
	\includegraphics [width=0.49\textwidth,clip, trim=2.5mm 67.7mm 2.5mm 7.5mm]{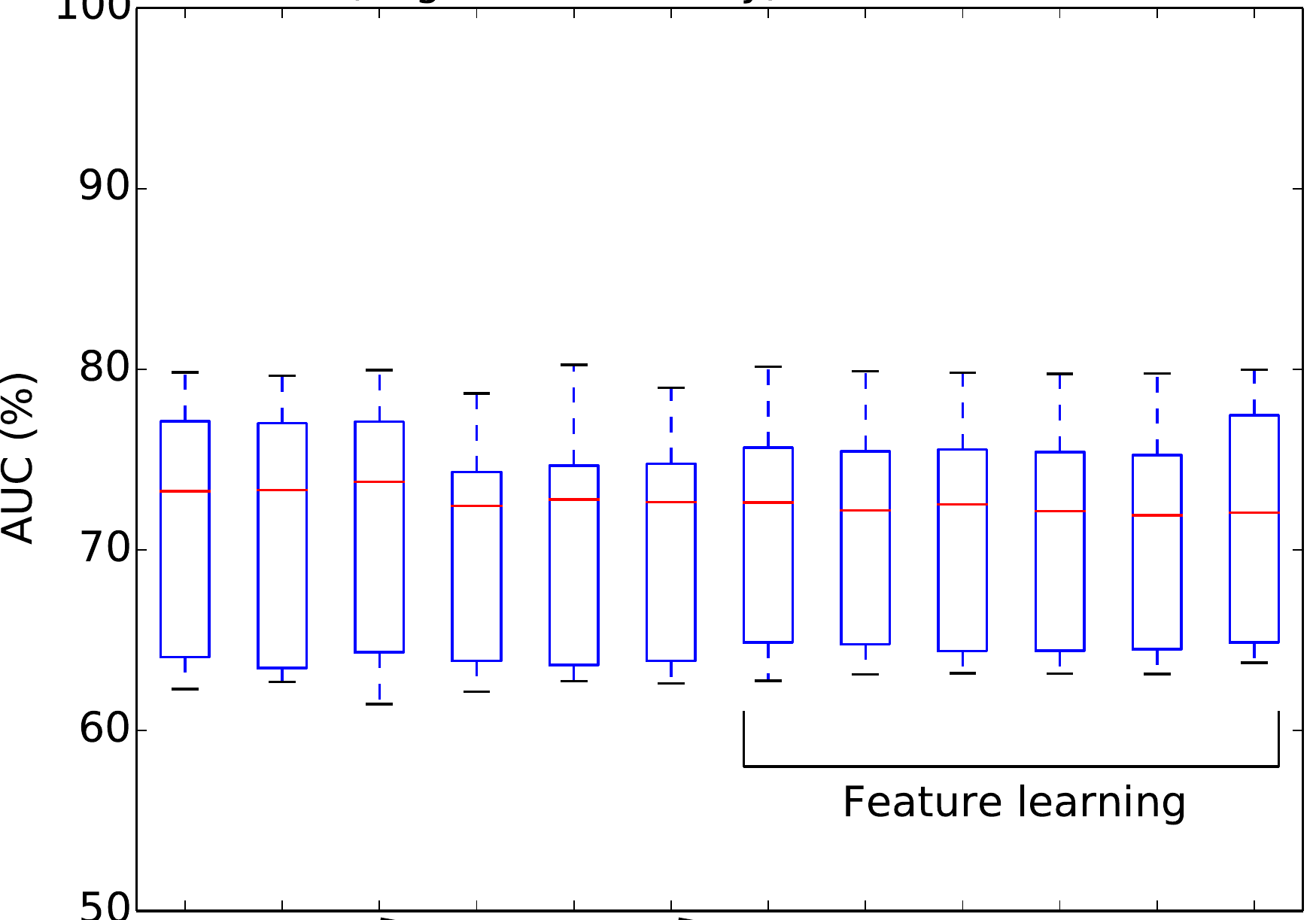}
	\includegraphics [width=0.49\textwidth,clip, trim=2.5mm 67.7mm 2.5mm 7.5mm]{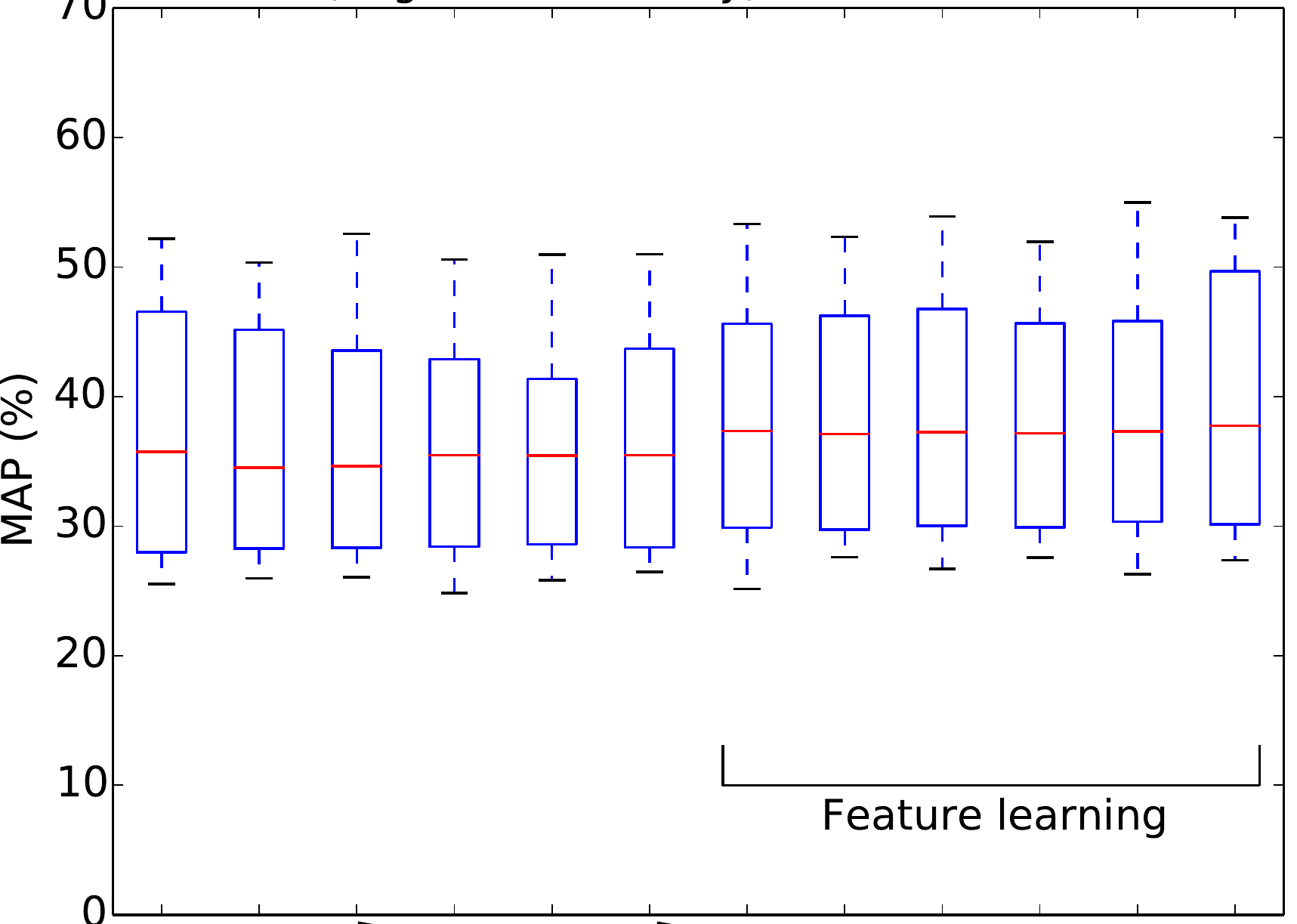}
	\\
	\includegraphics [width=0.49\textwidth,clip, trim=2.5mm 2.5mm 2.5mm 7.5mm]{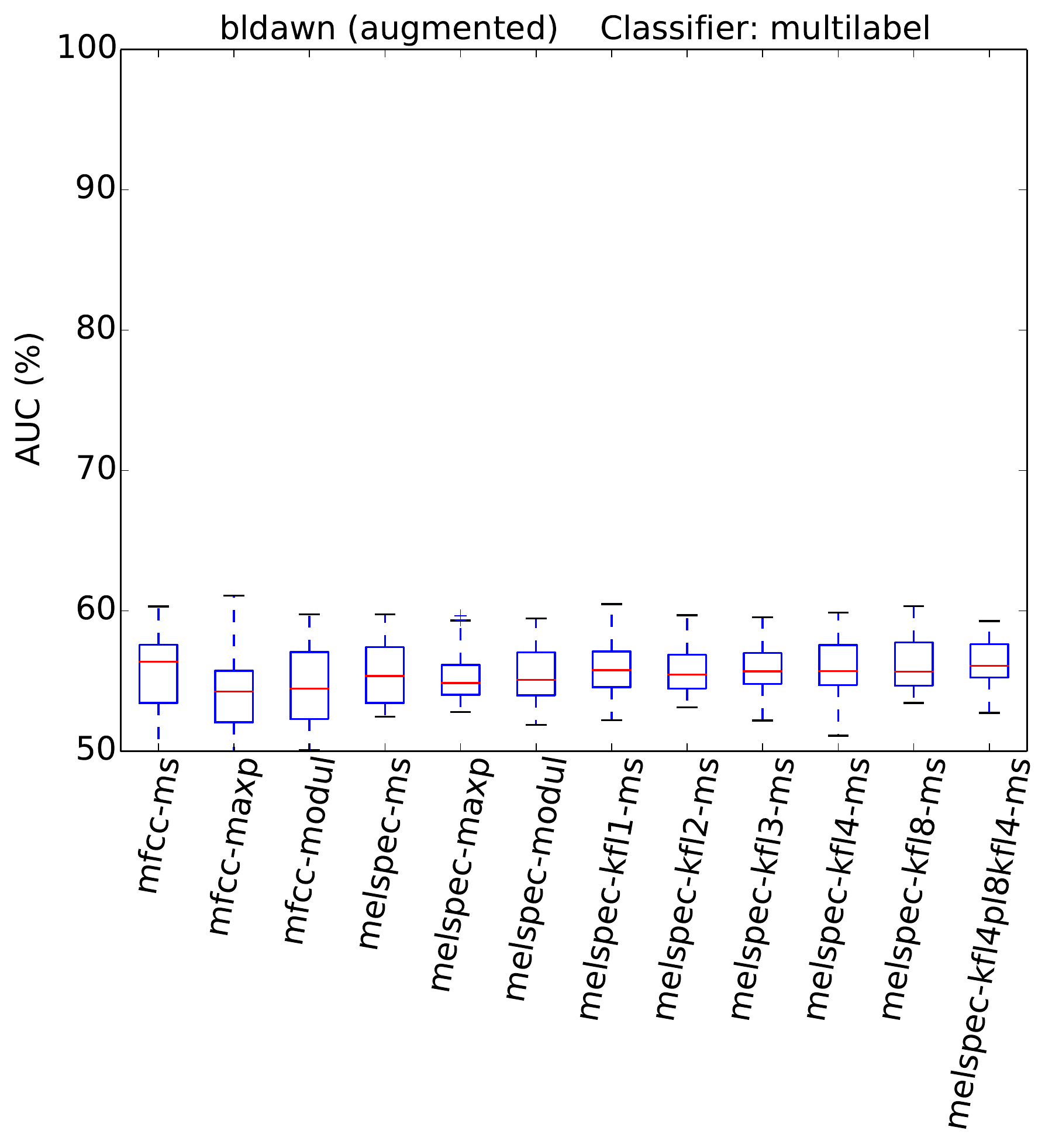}
	\includegraphics [width=0.49\textwidth,clip, trim=2.5mm 2.5mm 2.5mm 7.5mm]{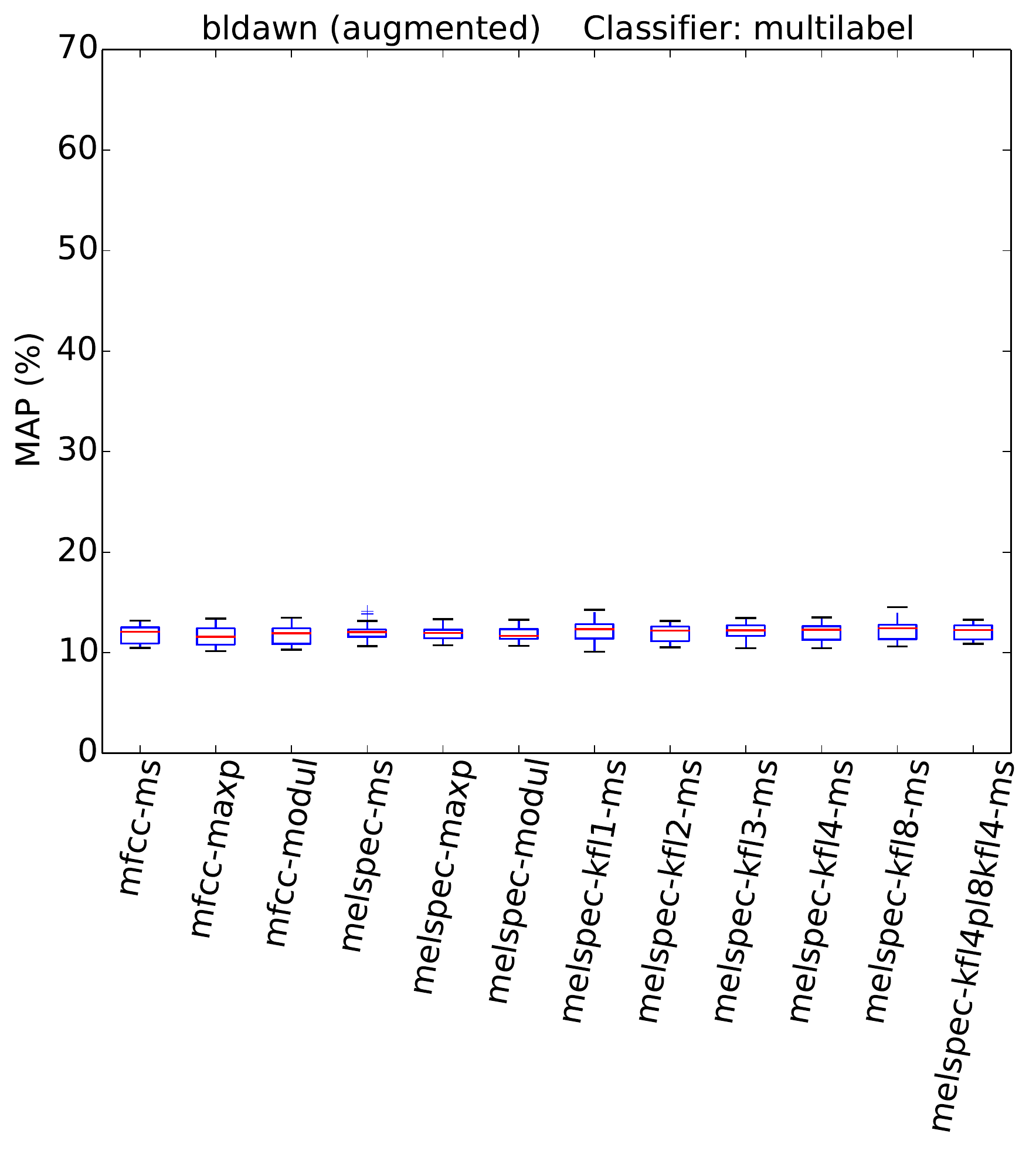}
	\end{center}
	\caption{As Fig. \ref{fig:resultsfeats}, for the \textit{bldawn} dataset, except that the input data for training is either: %
(a) a different dataset, \textit{xccoverbl} (cross-condition scenario; upper); %
(b) the union of the two datasets for feature learning, but \textit{bldawn} alone for training (middle); %
(c) the union of the two datasets for feature learning and for training (lower); %
}
	\label{fig:resultsfeatblxc}
\end{figure}

Our follow-up data expansion tests and cross-condition test failed to improve performance for \textit{bldawn} (Fig. \ref{fig:resultsfeatblxc}).
Adding the \textit{xccoverbl} data to the feature learning step made little difference,
giving a slight but insignificant boost to the two-layer model.
This tells us firstly that the dataset already contained enough audio for feature-learning to operate satisfactorily,
and secondly that the audio from \textit{xccoverbl} is similar enough in kind that its use is no detriment.
However, the story is quite different for the cases in which we then included \textit{xccoverbl} in the training step.
With or without feature learning, using \textit{xccoverbl} to provide additional training data for \textit{bldawn} acted as a distractor in this particular classification setup, and performed uniformly poorly.
Note that the data expansion procedure augmented the audio data size by around 60\%,
but augmented the number of annotations even more substantially (since \textit{xccoverbl} contains more individual items than \textit{bldawn}),
and so the classifier may have been led to accommodate the single-label data better than the dawn chorus annotations.
We diagnosed the problem by looking at the classification quality that the classifiers attained on their \textit{bldawn} training examples:
in this case the quality was poor (AUC $<60\%$),
confirming that the single-label \textit{xccoverbl} data had acted as distractors rather than additional educational examples.

\begin{figure}[tp]
	\begin{center}
	\includegraphics [width=0.49\textwidth,clip, trim=2.5mm 2.5mm 2.5mm 2.5mm]{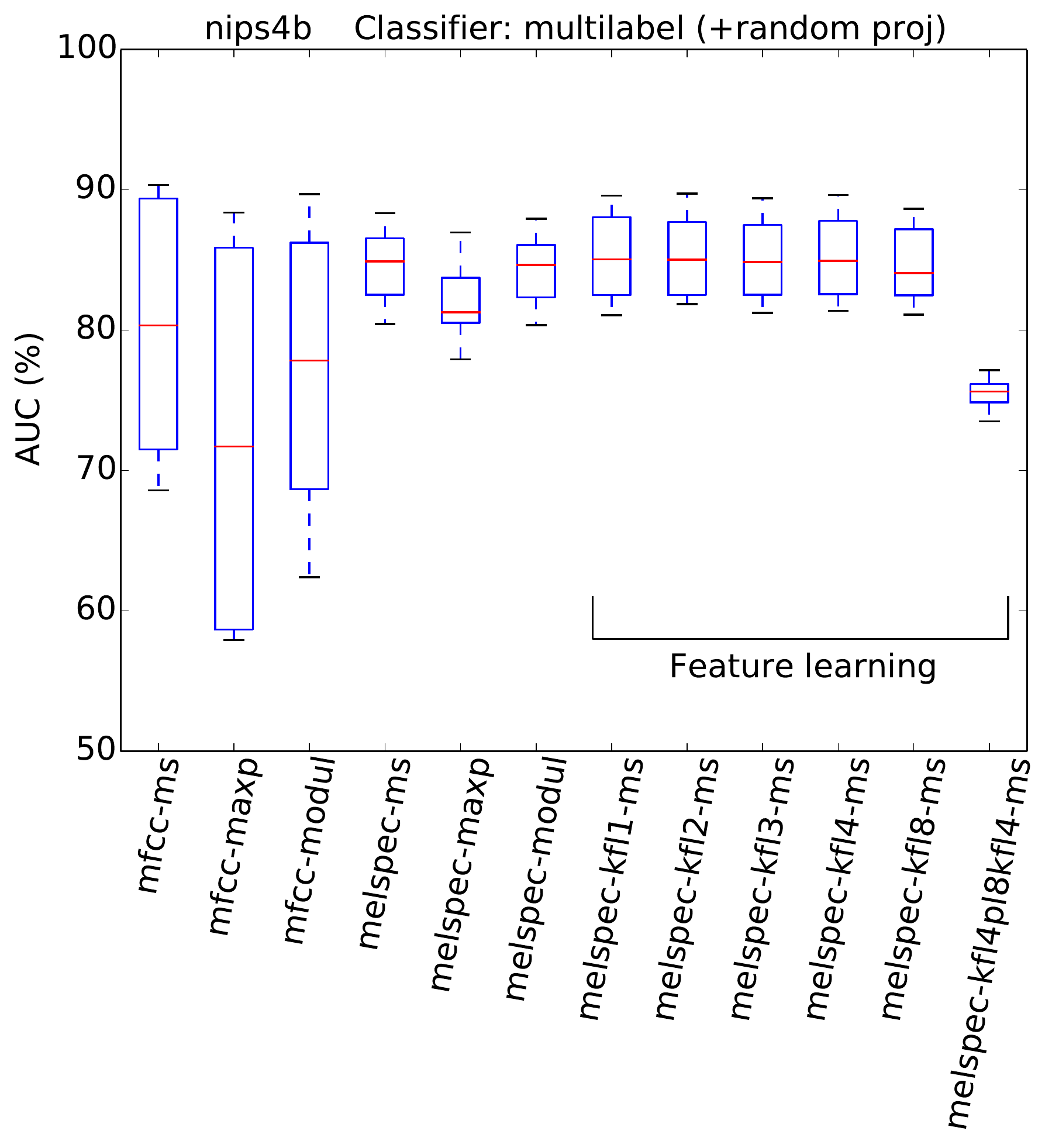}
	\includegraphics [width=0.49\textwidth,clip, trim=2.5mm 2.5mm 2.5mm 2.5mm]{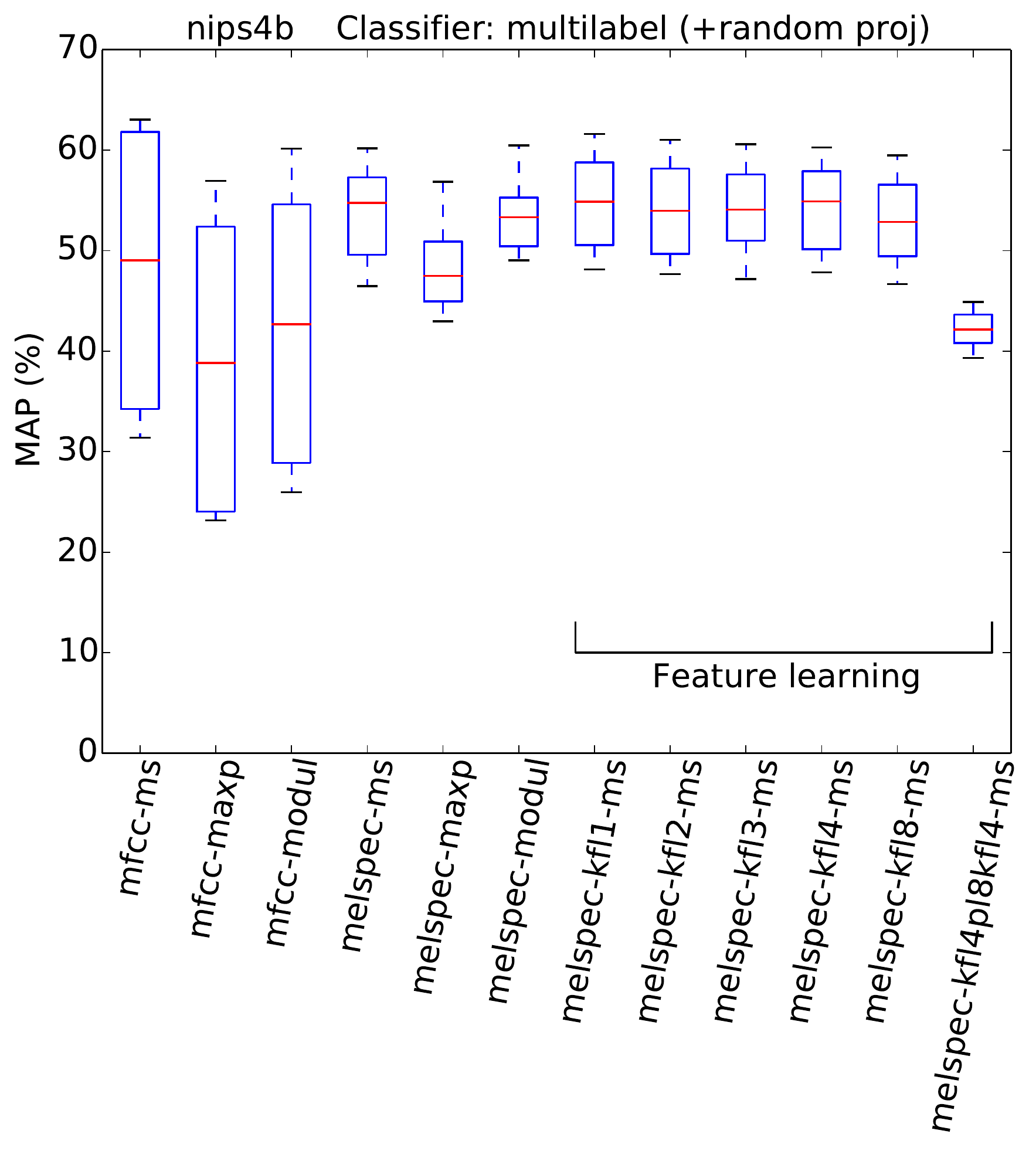}
	\end{center}
	\caption{As Fig. \ref{fig:resultsfeats}, for the \textit{nips4b} dataset, except that the feature dimensionality is standardised before train/test by the application of a random projection.}
	\label{fig:resultsfeatrfrpall}
\end{figure}

Turning to the \textit{nips4b} dataset to explore the effect of feature dimensionality,
the relative performance of the different feature types was broadly preserved even after dimensionality was standardised by random projection
(Fig. \ref{fig:resultsfeatrfrpall}).
The general effect of the random projection was a modest improvement for low-dimensional features,
and a modest impairment for high-dimensional (learned) features,
but not to the extent of changing the ordering of performance.
This suggests that high dimensionality is a small,
though non-zero, part of what lends the learned features their power.
The overall effect estimated by the GLM for the random-projection modification was a small but significant impairment ($-0.07$).

\begin{table}[t]
\resizebox{\textwidth}{!}{
\begin{tabular}{ l | r r }
	System variant submitted	&	Cross-validated MAP (\%)	& Final official MAP (\%)  \\
	\hline
	\texttt{melspec-kfl3-ms}, noise red., binary relevance	&	        30.56	&	        36.9	\\
	Average from 12 single-layer models	&	        32.73	&	        38.9	\\
	\texttt{melspec-kfl4pl8kfl4-ms}, noise red., binary relevance	&	\textbf{35.31}	&	\textbf{42.9}	\\
	Average from 16 single- and double-layer models	&	        35.07	&	        41.4	\\
\end{tabular}
}
\caption{Summary of MAP scores attained by our system in the public LifeCLEF 2014 Bird Identification Task \citep{birdclef2014}. %
The first column lists scores attained locally in our two-fold \textit{lifeclef2014} split. %
The second column lists scores evaluated officially, using a classifier(s) trained across the entire training set. %
}
\label{tbl:lifeclefofficial}
\end{table}

We can compare our results against those obtained recently by others.
A formal comparison was conducted in Spring 2014 when we submitted decisions from our system to the LifeCLEF 2014 bird identification challenge
\citep{birdclef2014}.
In that evaluation, our system attained by far the strongest audio-only classification results,
with a MAP peaking at 42.9\% (Table \ref{tbl:lifeclefofficial}).
(Only one system outperformed ours, peaking at 51.1\% in a variant of the challenge which provided additional metadata as well as audio.)
We submitted the outputs from individual models, as well as model-averaging runs using the simple mean of outputs from multiple models.
Notably, the strongest classification both in our own tests and the official evaluation was attained not by model averaging,
but by a single model based on two-layer feature learning.
Also notable is that our official scores, which were trained and tested on larger data subsets,
were substantially higher than our crossvalidated scores,
corroborating our observation that the method works particularly well at high data volumes.

Considering the \textit{nips4b} dataset,
the peak result from our main tests reached a crossvalidated AUC of 89.8\%.
In the actual NIPS4B contest (conducted before our current approach was developed), the winning result attained 91.8\%;
\cite{Potamitis:2014}, developing further a model submitted to the contest, reports a peak AUC of 91.7\%.
Our results are thus slightly behind these, although note that these other reported results use the full public-and-private datasets, without crossvalidation,
whereas we restricted ourselves only to the data that were fully public
and divided this public data into two crossvalidation folds, so the comparison is not strict.

%\subsection{Run times}

\begin{figure}[tp]
	\begin{center}
	\includegraphics [width=0.65\textwidth,clip, trim=0mm 0mm 0mm 0mm]{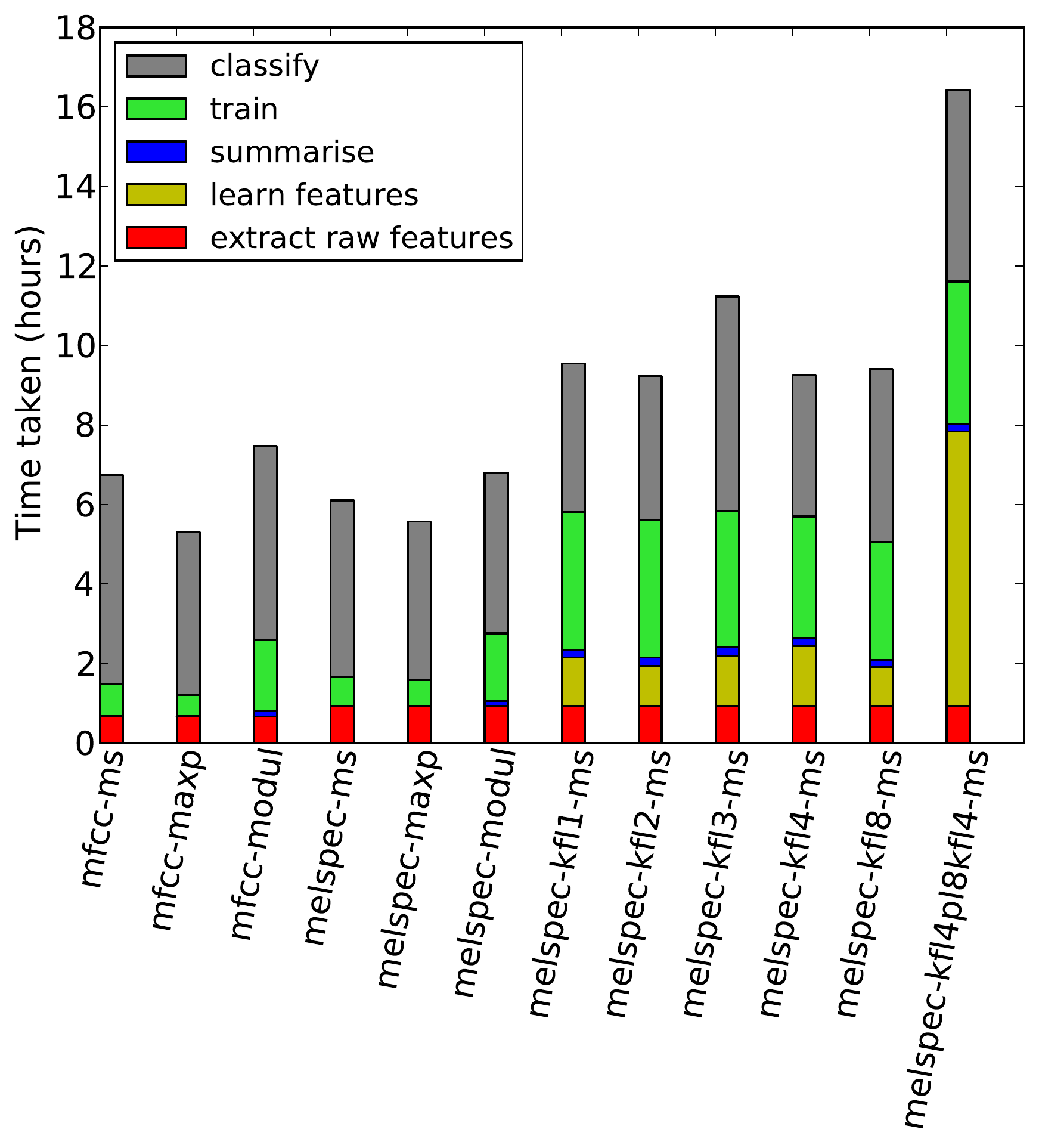}
	\end{center}
	\caption{Times taken for each step in the process, for the \textit{lifeclef2014} dataset. %
Note that these are heuristic ``wallclock'' times measured on processes across two compute servers, %
and disk read/write processes (to store state) took non-trivial time in each step. %
Each measurement is averaged across the two folds and across two settings (noise reduction on/off) across the runs using the multilabel classifier and no decision-pooling. %
}
	\label{fig:timestaken}
\end{figure}

We measured the total time taken for each step in our workflow,
to determine the approximate computational load for the steps (Fig. \ref{fig:timestaken}).
The timings are approximate---in particular because our code was modularised to save/load state on disk between each process,
which impacted particularly on the ``classify'' step which loaded large random forest settings from disk before processing.
Single-layer feature learning was efficient, taking a similar amount of time as did the initial feature extraction.
Double-layer feature learning took more than double this, because of the two layers as well as performing max-pooling downsampling.
Training the random forest classifier took longer on the learned features due to the higher dimensionality.
However, once the system was trained, the time taken to classify new data was the same across all configurations.

%%%%%%%%%%%%%%%%%%%%%%%%%%%%%%%%%%%%%%%%%%%%%%%
\section{Discussion}
\label{sec:disc}

In order to be of use for applications in ecology and archival,
automatic bird species recognition from sound must work across large data volumes,
across large numbers of potential species,
and on data with a realistic level of noise and variation.
Our experiments have demonstrated that very strong results can be achieved in exactly these cases
by supplementing a classification workflow with unsupervised feature learning.
We have here used a random forest classifier, but unsupervised feature learning operates without any knowledge
of the classifier or even the training labels, so we can expect this finding to apply in other classification systems (cf.\ \cite{Erhan:2010}).
The procedure requires large data volumes in order for benefits to be apparent,
as indicated by the lack of improvement on our \textit{bldawn} dataset,
and by the failure of two-layer feature learning on the \textit{nips4b} dataset.
However, the use of single-layer feature learning creates a classifier that is equivalent to or better than manually-designed features in all our tests.
There were very few differences in performance between our different versions of feature learning.
One difference is that two-layer feature learning, while unsuccessful for \textit{nips4b},
led to the strongest performance for \textit{lifeclef2014} which is the largest dataset considered%
---largest by an order of magnitude in data volume, and by almost an order of magnitude in the number of possible species labels.
This confirms the recommendations of \cite{Coates:2012} about the synergy between feature learning and big data scales,
here for the case of ecological audio data.

However, note that a lesser but still substantial improvement over the baseline MFCC system can usually be attained simply by using the raw Mel spectral data as input rather than MFCCs.
One of the long-standing motivations for the MFCC transformation has been to reduce spectral data down to a lower dimensionality while hoping to preserve most of the implicit semantic information;
but as we have seen, the random forest classifier performs well with high-dimensional input,
and such data reduction is not necessary and often holds back classification performance.
Future investigators should consider using Mel spectra as a baseline, rather than MFCCs as is common at present.

The lack of improvement on the \textit{bldawn} dataset is notable,
along with the low-quality results obtained by simply using a different dataset for training, or augmenting the data with an additional dataset.
The availability of audio data is not the issue here, as the dataset is second-largest by audio volume,
although the availability of training annotations may be crucial, since the dataset is the smallest by an order of magnitude
in terms of individual labelled items.
Augmenting the data with \textit{xccoverbl} increased the amount of training annotations,
but its failure to boost performance may have been due to differences in kind,
since it consisted of single-label recordings of individual birds rather than multilabel recordings of dawn chorus soundscapes.
The issue is not due to differences in recording conditions (such as microphone type or recording level):
the audio went through various levels of standardisation in our workflow,
and caused no problems when added to the feature-learning step only.
We emphasise that the audio and metadata of the \textit{bldawn} dataset comes directly from a sound archive collection,
and the long-form recordings with sparse annotation are exactly the format of a large portion of the holdings in sound archives.
Our results (up to around 80\% AUC) correspond to a classifier that could provide useful semi-automation of archive labelling
(e.g.\ suggested labels to a human annotator) but not yet a fully automatic process in that case.
This outcome thus reinforces the importance of collecting and publishing annotated audio datasets which fit the intended application.
Public collections such as Xeno Canto are highly valuable, but their data do not provide the basis to solve all tasks in species recognition,
let alone the other automated tasks (censusing, individual recogntion) that we may wish to perform automatically from audio data.

\begin{figure}[tp]
	\begin{center}
	\includegraphics [width=0.95\textwidth,clip, trim=15mm 20mm 40mm 0mm]{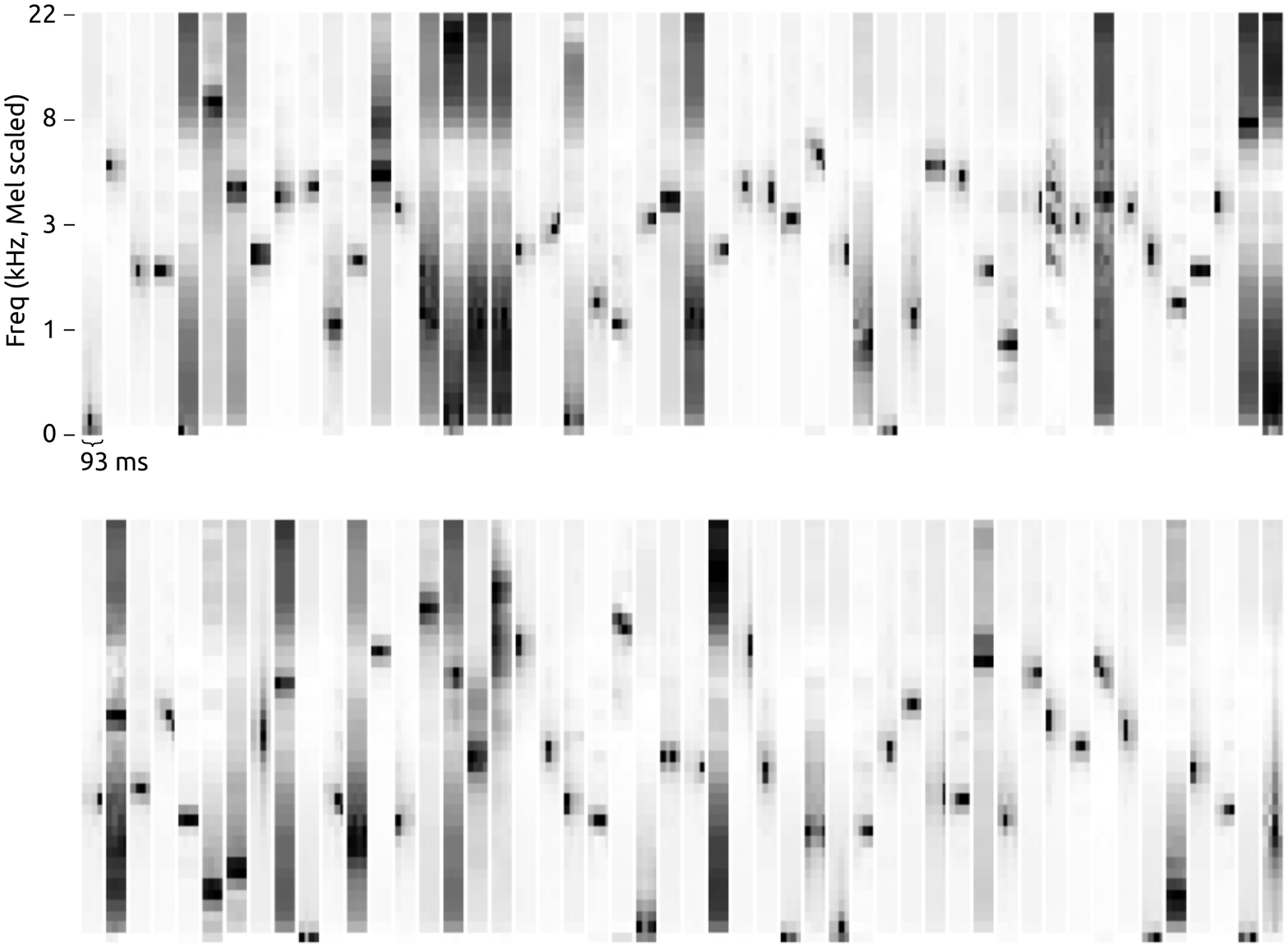}
	\end{center}
	\caption{A random subset of the spectrotemporal bases estimated during one run of feature learning, in this case using 4 frames per base and the \textit{lifeclef2014} dataset. %
Each base is visualised as a brief spectrogram excerpt, with dark indicating high values. %
The frequency axis is nonlinearly (Mel) scaled. %
}
	\label{fig:basesexample}
\end{figure}

\begin{figure}[tp]
	\begin{center}
	\includegraphics [width=0.65\textwidth,clip, trim=35mm 15mm 60mm 10mm]{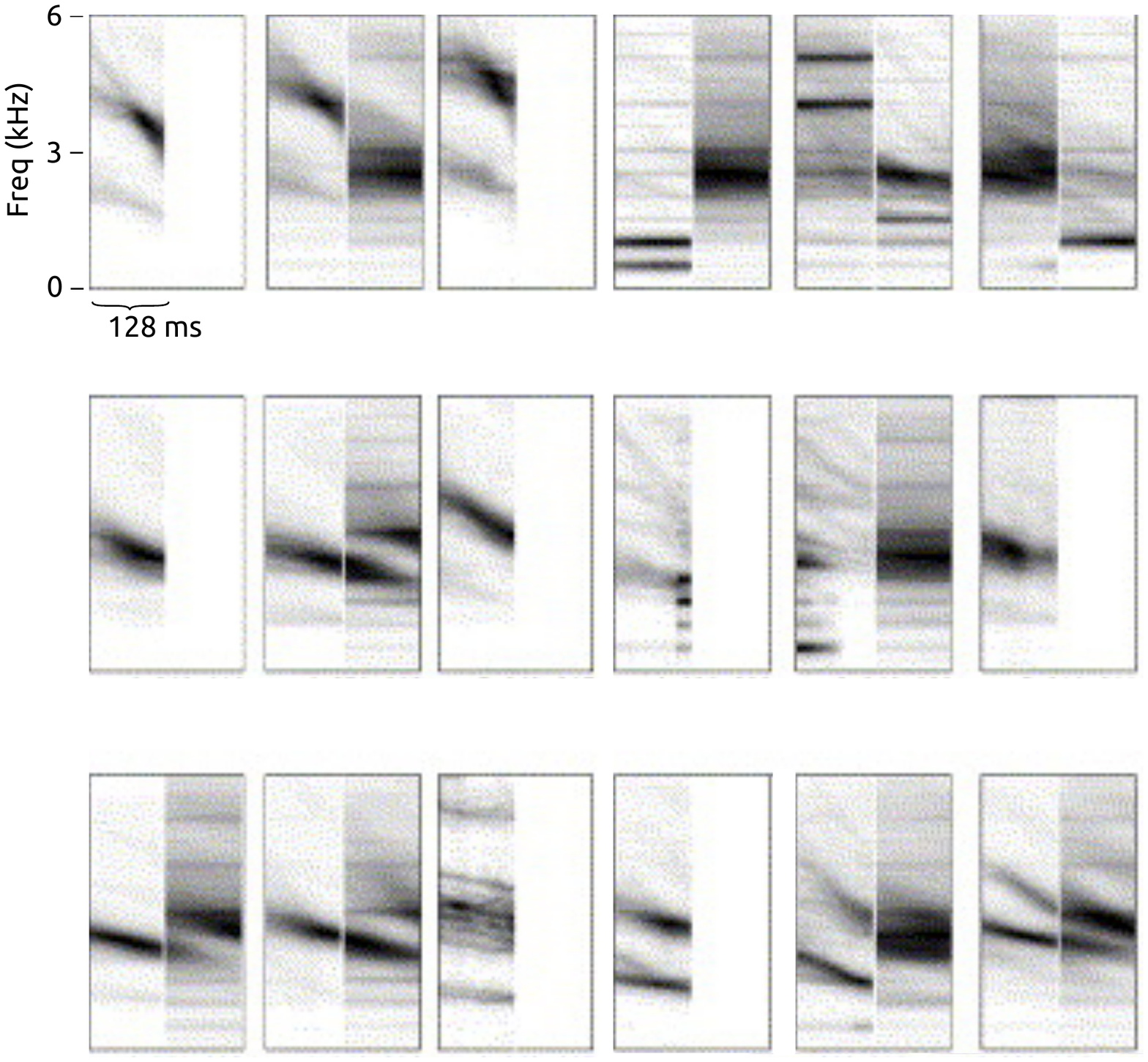}
	\end{center}
	\caption{Spectro-temporal receptive fields (STRFs) measured from individual neurons in auditory field L of starling. %
Adapted from \cite{Hausberger:2000} for comparison with Fig. \ref{fig:basesexample}. %
Each image shows the spectro-temporal patterns that correlate with excitation (left) and inhibition (right) of a single neuron. %
The frequency axis is linearly scaled. %
}
	\label{fig:hausbergerfig4}
\end{figure}

Our feature-based approach to classification is not the only approach.
Template-based methods have some history in the literature, with the main issue of concern being how to match a limited set of templates against the unbounded natural variation in bird sound realisations, in particular the dramatic temporal variability.
One technique to compensate for temporal variability is dynamic time warping (DTW)
\citep{Anderson:1996}%jasa, Template-based automatic recognition of birdsong syllables from continuous recordings
\citep{Ito:1999}% jasa, Dynamic programming matching as a simulation of budgerigar contact-call discrimination
. Recent methods which performed very strongly in the SABIOD-organised contests used templates without any time-warping considerations,
making use of a large number of statistics derived from the match between a template and an example (not using just the closeness-of-match)
\citep{Fodor:2013}% The Ninth Annual {MLSP} Competition: First place
. Other recent methods use templates and DTW but deployed within a kernel-based distance measure,
again going beyond a simple one-to-one match \citep{Damoulas:2010}.

In light of these other perspectives, we note an analogy with our learned features.
In the workflow considered here, the new representation is calculated by taking the dot-product between the input data and each base
(such as those in Fig. \ref{fig:basesexample}), as given in \eqref{eqn:dotprodmultiframe}.
The form of \eqref{eqn:dotprodmultiframe} is the same mathematically as spectro-temporal cross-correlation,
where the $b_j$ would be thought of more traditionally as ``templates''.
Our features are thus equivalent to the output of an unusual kind of template matching by cross-correlation,
where the ``templates'' are not indivudal audio excerpts but generalisations of features found broadly across audio excerpts,
and are also of a fixed short duration (shorter than many song syllables, though long enough to encompass many calls).

A question that arises from this perspective is whether our approach should use longer series of frames, long enough to encompass many types of song syllable entirely.
In our tests we found no notable tendency for improved recognition as we increased the number of frames from one to eight,
and we also saw many temporally-compact bases learnt (Fig. \ref{fig:basesexample}),
so we do not believe lengthening the bases is the route to best performance.
Further, the advantage of using relatively short durations is that the feature learning method learns \textit{components} of bird vocalisations rather than over-specific whole units.
These components may co-occur in a wide variety of bird sounds, in temporally-flexible orders,
conferring a combinatorial benefit of broad expressivity.
Our two-layer feature learning provides a further level of abstraction over temporal combinations of energy patterns,
which is perhaps part of its advantage when applied to our largest dataset.
We have not explicitly tested our method in comparison to template-based approaches;
the relative merits of such approaches will become clear in further research.

The bases shown in Fig. \ref{fig:basesexample} also bear some likeness with the spectro-temporal receptive fields (STRFs) measured from neurons found in the early auditory system of songbirds
(e.g.\ Fig. \ref{fig:hausbergerfig4}, adapted from \cite{Hausberger:2000})% Neuronal bases of categorization in starling song   ---- this has the Fig 4 that I like
. Broadly similar generalisations seem to emerge, including sensitivity to spectrally-compact stationary tones as well as up-chirps and down-chirps, and combinations of harmonic sounds.
We do not make strong claims from this likeness:
firstly because our method (spherical k-means) is a simple sparse feature learning method with no designed resemblance to neural processes involved in natural learning,
and secondly because STRFs show only a partial summary of the nonlinear response characteristics of neurons.
However, sparse feature learning in general is motivated by considerations of the informational and energetic constraints that may have influenced the evolution of neural mechanisms \citep{Olshausen:2004}.
\cite{Theunissen:2006} %  title={Auditory processing of vocal sounds in birds -- This is a great review. Short and readable too.  Explains how zenk expression relates to learning to recognise song sounds.
note that the sensitivities measured from neurons in the avian primary auditory forebrain generally relate not to entire song syllables, but to smaller units which may serve as building blocks for later processing.
Biological analogies are not a necessary factor in the power of machine learning methods,
but such hints from neurology suggest that the method we have used in this study fits within a paradigm that may be worth further exploration.

%%%%%%%%%%%%%%%%%%%%%%%%%%%%%%%%%%%%%%%%%%%%%%%
\section{Conclusions}
\label{sec:conc}

Current interest in automatic classification of bird sounds is motivated by the practical scientific need to label
large volumes of data coming from sources such as remote monitoring stations and sound archives.
Unsupervised feature learning is a simple and effective method to boost classification performance
by learning spectro-temporal regularities in the data.
It does not require training labels or any other side-information,
it can be used within any classification workflow,
and once trained it imposes negligible extra computational effort on the classifier.
In experiments it learns regularities in bird vocalisation data with similar qualities to the sensitivities of bird audition reported by others.

The principal practical issue with unsupervised feature learning is that it requires large data volumes to be effective,
as confirmed in our tests.
However, this exhibits a synergy with the large data volumes that are increasingly becoming standard.
For our largest dataset, feature learning led to classification performance up to 85.4\% AUC,
whereas without feature learning the performance peaked at 82.2\% for raw Mel spectra or 69.3\% for MFCCs.

In our tests, the choice of feature set made a much larger difference to classification performance than any of our other configuration choices
(such as the use of noise reduction, decision pooling, or binary relevance).
MFCCs cannot be recommended for bird species recognition since even the simple approach of using undigested Mel spectra dramatically outperforms them in most of our tests.
Although MFCCs have been widespread as baseline features, the Mel spectra themselves may be more appropriate for benchmarking.
Across our various tests in single-label and multilabel settings,
unsupervised feature learning together with a multilabel classifier achieved peak or near-peak classification quality.

This study, thanks to the large-scale data made available by others,
has demonstrated strong performance on bird sound classification is possible at very large scale, when the synergy between big data volumes and feature learning is exploited.
However, automatic classification is not yet trivial across all domains, as demonstrated by the lack of improvement on our \textit{bldawn} dataset of dawn chorus recordings.
The research community will benefit most from the creation/publication of large bird audio collections,
labelled or at least part-labelled, and published under open data licences.

%%%%%%%%%%%%%%%%%%%%%%%%%%%%%%%%%%%%%%%%%%%%%%%
\section*{Acknowledgments}

We would like to thank the people and projects which made available the data used for this research:
the Xeno Canto website and its many volunteer contributors;
the SABIOD research project and the Biotope society;
the British Library Sound Archive and its contributors, and curator Cheryl Tipp.

%%%%%%%%%%%%%%%%%%%%%%%%%%%%%%%%%%%%%%%%%%%%%%%
\section*{Funding sources}

This work was supported by EPSRC Leadership Fellowship EP/G007144/1 and EPSRC Early Career Fellowship EP/L020505/1.

%%%%%%%%%%%%%%%%%%%%%%%%%%%%%%%%%%%%%%%%%%%%%%%
\section*{Data availability}

\begin{itemize}
\item	The \textit{xccoverbl} data is archived at \url{https://archive.org/details/xccoverbl_2014} --- it is composed of sound files sourced from \url{http://www.xeno-canto.org/}. The sound file authors and other metadata are listed in a CSV file included in the dataset.

\item	The \textit{bldawn} dataset is available on request from the British Library Sound Archive (BLSA).
Our machine-readable version of the species metadata is downloadable from Figshare,
and lists the file identifiers corresponding to the BLSA records:
\url{http://dx.doi.org/10.6084/m9.figshare.1024549}

\item	The \textit{nips4b} dataset is downloadable from the SABIOD/nips4b website:
\url{http://sabiod.univ-tln.fr/nips4b/challenge1.html}

\item	The \textit{lifeclef2014} dataset is available from the Lifeclef website:
\url{http://www.imageclef.org/2014/lifeclef/bird}

\end{itemize}

\bibliographystyle{plainnat}
\bibliography{../refs}

\begin{thebibliography}{42}
\providecommand{\natexlab}[1]{#1}
\providecommand{\url}[1]{\texttt{#1}}
\expandafter\ifx\csname urlstyle\endcsname\relax
  \providecommand{\doi}[1]{doi: #1}\else
  \providecommand{\doi}{doi: \begingroup \urlstyle{rm}\Url}\fi

\bibitem[Acevedo et~al.(2009)Acevedo, Corrada-Bravo, Corrada-Bravo,
  Villanueva-Rivera, and Aide]{Acevedo:2009}
M.~A. Acevedo, C.~J. Corrada-Bravo, H.~Corrada-Bravo, L.~J. Villanueva-Rivera,
  and T.~M. Aide.
\newblock Automated classification of bird and amphibian calls using machine
  learning: A comparison of methods.
\newblock \emph{Ecological Informatics}, 4\penalty0 (4):\penalty0 206--214,
  2009.
\newblock \doi{10.1016/j.ecoinf.2009.06.005}.

\bibitem[Aide et~al.(2013)Aide, Corrada-Bravo, Campos-Cerqueira, Milan, Vega,
  and Alvarez]{Aide:2013}
T.~M. Aide, C.~Corrada-Bravo, M.~Campos-Cerqueira, C.~Milan, G.~Vega, and
  R.~Alvarez.
\newblock Real-time bioacoustics monitoring and automated species
  identification.
\newblock \emph{PeerJ}, 1:\penalty0 e103, 2013.
\newblock \doi{10.7717/peerj.103}.

\bibitem[Anderson et~al.(1996)Anderson, Dave, and Margoliash]{Anderson:1996}
S.~E. Anderson, A.~S. Dave, and D.~Margoliash.
\newblock Template-based automatic recognition of birdsong syllables from
  continuous recordings.
\newblock \emph{Journal of the Acoustical Society of America}, 100\penalty0 (2,
  Part 1):\penalty0 1209--1219, Aug 1996.
\newblock ISSN {0001-4966}.

\bibitem[Ballmer et~al.(2013)]{BTO:2013}
D.~Ballmer et~al.
\newblock \emph{BTO Bird atlas 2007--11: The breeding and wintering birds of
  Britain and Ireland}.
\newblock 2013.

\bibitem[Bates et~al.(2014)Bates, Maechler, Bolker, and Walker]{Bates:2014}
D.~Bates, M.~Maechler, B.~Bolker, and S.~Walker.
\newblock \emph{lme4: Linear mixed-effects models using Eigen and S4}, 2014.
\newblock URL \url{http://CRAN.R-project.org/package=lme4}.
\newblock R package version 1.0-6.

\bibitem[Bengio et~al.(2013)Bengio, Courville, and Vincent]{Bengio:2013}
Y.~Bengio, A.~Courville, and P.~Vincent.
\newblock Representation learning: A review and new perspectives.
\newblock \emph{IEEE Trans. Pattern Anal. Mach. Intell.}, 35\penalty0
  (8):\penalty0 1798–1828, 2013.
\newblock ISSN 2160-9292.
\newblock \doi{10.1109/tpami.2013.50}.

\bibitem[Breiman(2001)]{Breiman:2001}
L.~Breiman.
\newblock Random forests.
\newblock \emph{Machine Learning}, 45\penalty0 (1):\penalty0 5--32, 2001.
\newblock \doi{10.1023/A:1010933404324}.

\bibitem[Briggs et~al.(2009)Briggs, Raich, and Fern]{Briggs:2009}
F.~Briggs, R.~Raich, and X.~Z. Fern.
\newblock Audio classification of bird species: A statistical manifold
  approach.
\newblock In \emph{Proceedings of the Ninth {IEEE} International Conference on
  Data Mining}, pages 51--60, Dec 2009.
\newblock \doi{10.1109/ICDM.2009.65}.

\bibitem[Briggs et~al.(2012)Briggs, Lakshminarayanan, Neal, Fern, Raich,
  Hadley, Hadley, and Betts]{Briggs:2012}
F.~Briggs, B.~Lakshminarayanan, L.~Neal, X.~Z. Fern, R.~Raich, S.~J.~K. Hadley,
  A.~S. Hadley, and M.~G. Betts.
\newblock Acoustic classification of multiple simultaneous bird species: A
  multi-instance multi-label approach.
\newblock \emph{Journal of the Acoustical Society of America}, 131:\penalty0
  4640--4650, 2012.
\newblock \doi{10.1121/1.4707424}.

\bibitem[Caruana and Niculescu-Mizil(2006)]{Caruana:2006}
R.~Caruana and A.~Niculescu-Mizil.
\newblock An empirical comparison of supervised learning algorithms.
\newblock In \emph{Proceedings of the 23rd international conference on Machine
  learning}, pages 161--168. ACM, 2006.
\newblock \doi{10.1145/1143844.1143865}.

\bibitem[Coates and Ng(2012)]{Coates:2012}
A.~Coates and A.~Y. Ng.
\newblock Learning feature representations with k-means.
\newblock In \emph{Neural Networks: Tricks of the Trade}, pages 561--580.
  Springer, 2012.
\newblock \doi{10.1007/978-3-642-35289-8_30}.

\bibitem[Damoulas et~al.(2010)Damoulas, Henry, Farnsworth, Lanzone, and
  Gomes]{Damoulas:2010}
T.~Damoulas, S.~Henry, A.~Farnsworth, M.~Lanzone, and C.~Gomes.
\newblock Bayesian classification of flight calls with a novel dynamic time
  warping kernel.
\newblock In \emph{Machine Learning and Applications (ICMLA), 2010 Ninth
  International Conference on}, pages 424--429, 2010.
\newblock \doi{10.1109/ICMLA.2010.69}.

\bibitem[Davis and Mermelstein(1980)]{Davis:1980}
S.~B. Davis and P.~Mermelstein.
\newblock Comparison of parametric representations for monosyllabic word
  recognition in continuously spoken sentences.
\newblock In \emph{{IEEE} Transactions on Acoustics, Speech, and Signal
  Processing}, pages 357--366, 1980.

\bibitem[Dieleman and Schrauwen(2013)]{Dieleman:2013}
S.~Dieleman and B.~Schrauwen.
\newblock Multiscale approaches to music audio feature learning.
\newblock In \emph{Proceedings of the International Conference on Music
  Information Retrieval (ISMIR 2013)}, 2013.

\bibitem[Digby et~al.(2013)Digby, Towsey, Bell, and Teal]{Digby:2013}
A.~Digby, M.~Towsey, B.~D. Bell, and P.~D. Teal.
\newblock A practical comparison of manual and autonomous methods for acoustic
  monitoring.
\newblock \emph{Methods in Ecology and Evolution}, 4\penalty0 (7):\penalty0
  675--683, 2013.
\newblock \doi{10.1111/2041-210X.12060}.

\bibitem[Erhan et~al.(2010)Erhan, Bengio, Courville, Manzagol, Vincent, and
  Bengio]{Erhan:2010}
D.~Erhan, Y.~Bengio, A.~Courville, P.-A. Manzagol, P.~Vincent, and S.~Bengio.
\newblock Why does unsupervised pre-training help deep learning?
\newblock \emph{Journal of Machine Learning Research}, 11:\penalty0 625--660,
  2010.

\bibitem[Fawcett(2006)]{Fawcett:2006}
T.~Fawcett.
\newblock An introduction to {ROC} analysis.
\newblock \emph{Pattern Recognition Letters}, 27\penalty0 (8):\penalty0
  861--874, 2006.
\newblock \doi{10.1016/j.patrec.2005.10.010}.

\bibitem[Fodor(2013)]{Fodor:2013}
G.~Fodor.
\newblock The ninth annual {MLSP} competition: First place.
\newblock In \emph{Proceedings of the International Conference on Machine
  Learning for Signal Processing (MLSP 2013)}, page 2pp. IEEE, 2013.
\newblock \doi{10.1109/MLSP.2013.6661932}.

\bibitem[Fox(2008)]{Fox:2008}
E.~J.~S. Fox.
\newblock \emph{Call-independent identification in birds}.
\newblock PhD thesis, University of Western Australia, 2008.
\newblock URL \url{http://theses.library.uwa.edu.au/adt-WU2008.0218}.

\bibitem[Glotin et~al.(2013)Glotin, LeCun, Arti\`eres, Mallat, Tchernichovski,
  and Halkias]{Glotin:2013}
H.~Glotin, Y.~LeCun, T.~Arti\`eres, S.~Mallat, O.~Tchernichovski, and
  X.~Halkias, editors.
\newblock \emph{Neural Information Processing Scaled for Bioacoustics, from
  Neurons to Big Data}, USA, 2013.
\newblock URL \url{http://sabiod.org/NIPS4B2013_book.pdf}.

\bibitem[Go{\"e}au et~al.(2014)Go{\"e}au, Glotin, Vellinga, and
  Rauber]{birdclef2014}
H.~Go{\"e}au, H.~Glotin, W.-P. Vellinga, and A.~Rauber.
\newblock {LifeCLEF} bird identification task 2014.
\newblock In \emph{CLEF working notes 2014}, 2014.

\bibitem[Hausberger et~al.(2000)Hausberger, Leppelsack, Richard, and
  Leppelsack]{Hausberger:2000}
M.~Hausberger, E.~Leppelsack, J.-P. Richard, and H.~J. Leppelsack.
\newblock Neuronal bases of categorization in starling song.
\newblock \emph{Behavioural Brain Research}, 114\penalty0 (1):\penalty0 89--95,
  2000.
\newblock \doi{10.1016/S0166-4328(00)00191-1}.

\bibitem[Ito and Mori(1999)]{Ito:1999}
K.~Ito and K.~Mori.
\newblock Dynamic programming matching as a simulation of budgerigar
  contact-call discrimination.
\newblock \emph{Journal of the Acoustical Society of America}, 105\penalty0
  (1):\penalty0 552--559, Jan 1999.
\newblock ISSN 0001-4966.
\newblock \doi{10.1121/1.424591}.

\bibitem[Jafari and Plumbley(2011)]{Jafari:2011}
M.~Jafari and M.~D. Plumbley.
\newblock Fast dictionary learning for sparse representations of speech
  signals.
\newblock \emph{IEEE Journal of Selected Topics in Signal Processing},
  5\penalty0 (5):\penalty0 1025--1031, 2011.
\newblock \doi{10.1109/JSTSP.2011.2157892}.

\bibitem[Laiolo(2010)]{Laiolo:2010}
P.~Laiolo.
\newblock The emerging significance of bioacoustics in animal species
  conservation.
\newblock \emph{Biological Conservation}, 143\penalty0 (7):\penalty0
  1635--1645, 2010.
\newblock \doi{10.1016/j.biocon.2010.03.025}.

\bibitem[Lakshminarayanan et~al.(2009)Lakshminarayanan, Raich, and
  Fern]{Lakshminarayanan:2009}
B.~Lakshminarayanan, R.~Raich, and X.~Fern.
\newblock A syllable-level probabilistic framework for bird species
  identification.
\newblock In \emph{Proceedings of the 2009 International Conference on Machine
  Learning and Applications}, pages 53--59, 2009.
\newblock \doi{10.1109/ICMLA.2009.79}.

\bibitem[Lee et~al.(2008)Lee, Han, and Chuang]{Lee:2008}
C.-H. Lee, C.-C. Han, and C.-C. Chuang.
\newblock Automatic classification of bird species from their sounds using
  two-dimensional cepstral coefficients.
\newblock \emph{{IEEE} Transactions on Audio and Speech and Language
  Processing}, 16\penalty0 (8):\penalty0 1541--1550, Nov 2008.
\newblock ISSN 1558-7916.
\newblock \doi{10.1109/TASL.2008.2005345}.

\bibitem[Lloyd(1982)]{Lloyd:1982}
S.~Lloyd.
\newblock Least squares quantization in {PCM}.
\newblock \emph{Information Theory, IEEE Transactions on}, 28\penalty0
  (2):\penalty0 129--137, 1982.
\newblock \doi{10.1109/TIT.1982.1056489}.

\bibitem[McFee(2012)]{McFee:2012}
B.~McFee.
\newblock \emph{More like this: Machine learning approaches to music
  similarity}.
\newblock PhD thesis, University of California, San Diego, 2012.
\newblock URL
  \url{http://cseweb.ucsd.edu/~bmcfee/papers/bmcfee_dissertation.pdf}.

\bibitem[McIlraith and Card(1997)]{McIlraith:1997}
A.~L. McIlraith and H.~C. Card.
\newblock Birdsong recognition using backpropagation and multivariate
  statistics.
\newblock \emph{IEEE Transactions on Signal Processing}, 45\penalty0
  (11):\penalty0 2740--2748, Nov 1997.
\newblock \doi{10.1109/78.650100}.

\bibitem[Olshausen and Field(2004)]{Olshausen:2004}
B.~A. Olshausen and D.~J. Field.
\newblock Sparse coding of sensory inputs.
\newblock \emph{Current Opinion in Neurobiology}, 14\penalty0 (4):\penalty0
  481--487, 2004.
\newblock \doi{10.1016/j.conb.2004.07.007}.

\bibitem[Pedregosa et~al.(2011)Pedregosa, Varoquaux, Gramfort, Michel, Thirion,
  Grisel, Blondel, Prettenhofer, Weiss, Dubourg, et~al.]{Pedregosa:2011}
F.~Pedregosa, G.~Varoquaux, A.~Gramfort, V.~Michel, B.~Thirion, O.~Grisel,
  M.~Blondel, P.~Prettenhofer, R.~Weiss, V.~Dubourg, et~al.
\newblock Scikit-learn: Machine learning in {P}ython.
\newblock \emph{Journal of Machine Learning Research}, 12:\penalty0 2825--2830,
  2011.

\bibitem[Potamitis(2014)]{Potamitis:2014}
I.~Potamitis.
\newblock Automatic classification of a taxon-rich community recorded in the
  wild.
\newblock \emph{PLOS ONE}, 9\penalty0 (5):\penalty0 e96936, May 2014.
\newblock \doi{10.1371/journal.pone.0096936}.

\bibitem[{R Core Team}(2012)]{R}
{R Core Team}.
\newblock \emph{R: A Language and Environment for Statistical Computing}.
\newblock R Foundation for Statistical Computing, Vienna, Austria, 2012.
\newblock URL \url{http://www.R-project.org/}.
\newblock {ISBN} 3-900051-07-0.

\bibitem[Ranft(2004)]{Ranft:2004}
R.~Ranft.
\newblock Natural sound archives: Past, present and future.
\newblock \emph{Anais da Academia Brasileira de Ci{\^e}ncias}, 76\penalty0
  (2):\penalty0 456--460, 2004.
\newblock \doi{10.1590/S0001-37652004000200041}.

\bibitem[Selin et~al.(2007)Selin, Turunen, and Tanttu]{Selin:2007}
A.~Selin, J.~Turunen, and J.~T. Tanttu.
\newblock Wavelets in recognition of bird sounds.
\newblock \emph{EURASIP Journal on Applied Signal Processing}, 2007\penalty0
  (1):\penalty0 141, 2007.
\newblock \doi{10.1155/2007/51806}.

\bibitem[Stowell and Plumbley(2010)]{Stowell:2010e}
D.~Stowell and M.~D. Plumbley.
\newblock Birdsong and {C4DM}: A survey of {UK} birdsong and machine
  recognition for music researchers.
\newblock Technical Report C4DM-TR-09-12, Centre for Digital Music, Queen Mary
  University of London, Aug 2010.
\newblock URL
  \url{http://c4dm.eecs.qmul.ac.uk/papers/2010/Stowell2010-C4DM-TR-09-12-birdsong.pdf}.

\bibitem[Stowell and Plumbley(2013)]{Stowell:2013n}
D.~Stowell and M.~D. Plumbley.
\newblock Feature design for multilabel bird song classification in noise
  (nips4b challenge).
\newblock In \emph{Proceedings of NIPS4b: Neural Information Processing Scaled
  for Bioacoustics, from Neurons to Big Data}, 2013.

\bibitem[Stowell and Plumbley(2014)]{Stowell:2014}
D.~Stowell and M.~D. Plumbley.
\newblock Large-scale analysis of frequency modulation in birdsong databases.
\newblock \emph{Submitted}, 2014.

\bibitem[Theunissen and Shaevitz(2006)]{Theunissen:2006}
F.~E. Theunissen and S.~S. Shaevitz.
\newblock Auditory processing of vocal sounds in birds.
\newblock \emph{Current Opinion in Neurobiology}, 16\penalty0 (4):\penalty0
  400--407, 2006.
\newblock \doi{10.1016/j.conb.2006.07.003}.

\bibitem[Tsoumakas et~al.(2010)Tsoumakas, Katakis, and
  Vlahavas]{Tsoumakas:2010}
G.~Tsoumakas, I.~Katakis, and I.~Vlahavas.
\newblock Mining multi-label data.
\newblock In \emph{Data mining and knowledge discovery handbook}, pages
  667--685. Springer, 2010.
\newblock \doi{10.1007/978-0-387-09823-4_34}.

\bibitem[Yue et~al.(2007)Yue, Finley, Radlinski, and Joachims]{Yue:2007}
Y.~Yue, T.~Finley, F.~Radlinski, and T.~Joachims.
\newblock A support vector method for optimizing average precision.
\newblock In \emph{Proceedings of the International Conference on Research and
  development in information retrieval (SIGIR 2007)}, pages 271--278. ACM,
  2007.
\newblock \doi{10.1145/1277741.1277790}.

\end{thebibliography}

\end{document}